\documentclass[conference]{IEEEtran}
\IEEEoverridecommandlockouts
\usepackage{graphicx}
\usepackage{todonotes}
\usepackage{color,soul}
\usepackage{url}
\usepackage{forest}
\usepackage{multicol}
\usepackage{cite}
\usepackage{amsmath}

\usepackage{amsmath,amssymb,amsfonts}
\usepackage{algorithmic}
\usepackage{textcomp}
\usepackage{multirow}
\usepackage{latexsym}
\usepackage{pifont}
\usepackage{subcaption}
\usepackage{lineno}
\usepackage{diagbox}
\usepackage{cleveref}
\usepackage[ruled,vlined, linesnumbered]{algorithm2e}

\usepackage{xcolor}
\def\BibTeX{{\rm B\kern-.05em{\sc i\kern-.025em b}\kern-.08em
    T\kern-.1667em\lower.7ex\hbox{E}\kern-.125emX}}
\begin{document}

\title{Graph-based Gossiping for Communication Efficiency in Decentralized Federated Learning\\

\thanks{* Corresponding author}
}

\author{\IEEEauthorblockN{Huong Nguyen}
\IEEEauthorblockA{
% \textit{Future Computing Group, UBICOMP} \\
\textit{University of Oulu}\\
Oulu, Finland\\
huong.nguyen@oulu.fi}
\and
\IEEEauthorblockN{Hong-Tri Nguyen}
\IEEEauthorblockA{
% \textit{dept. name of organization (of Aff.)} \\
\textit{Aalto University}\\
Espoo, Finland \\
 hong-tri.nguyen@aalto.fi}
\and
\IEEEauthorblockN{Praveen Kumar Donta}
\IEEEauthorblockA{
% \textit{dept. name of organization (of Aff.)}
\textit{Stockholm University}\\
Stockholm, Sweden \\
praveen.donta@dsv.su.se}
\and
\IEEEauthorblockN{Susanna Pirttikangas}
\IEEEauthorblockA{
% \textit{dept. name of organization (of Aff.)} \\
\textit{University of Oulu}\\
Oulu, Finland \\
susanna.pirttikangas@oulu.fi}
\and
\IEEEauthorblockN{Lauri Lovén\textsuperscript{*}}
\IEEEauthorblockA{
% \textit{Future Computing Group, UBICOMP} \\
\textit{University of Oulu}\\
Oulu, Finland \\
lauri.loven@oulu.fi}
}

\maketitle

\begin{abstract}
Federated learning has emerged as a privacy-preserving technique for collaborative model training across heterogeneously distributed silos. Yet, its reliance on a single central server introduces potential bottlenecks and risks of single-point failure. Decentralizing the server, often referred to as decentralized learning, addresses this problem by distributing the server’s role across nodes within the network. One drawback regarding this pure decentralization is it introduces communication inefficiencies, which arise from increased message exchanges in large-scale setups. However, existing proposed solutions often fail to simulate the real-world distributed and decentralized environment in their experiments, leading to unreliable performance evaluations and limited applicability in practice. Recognizing the lack from prior works, this work investigates the correlation between model size and network latency, a critical factor in optimizing decentralized learning communication. We propose a graph-based gossiping mechanism, where specifically, minimum spanning tree and graph coloring are used to optimize network structure and scheduling for efficient communication across various network topologies and message capacities. Our approach configures and manages subnetworks on real physical routers and devices and closely models real-world distributed setups. Experimental results demonstrate that our method significantly improves communication, compatible with different topologies and data sizes, reducing bandwidth and transfer time by up to circa 8 and 4.4 times, respectively, compared to naive flooding broadcasting methods.


\end{abstract}

\section{Introduction}
Collaborative learning~\cite{Liu_2024_CVPR} has emerged as a transformative paradigm in the field of machine learning, enabling distributed silos to collaboratively train models while preserving data privacy and towards the global model's convergence.
Federated Learning (FL)~\cite{McMahanMRHA17} known as Centralized Federated Learning (CFL) servers serve as central aggregators for collecting models from distributed participants, doing aggregation, and acting as the focal point of knowledge exchange. Consequently, the crux of the problem with large-scale CFLs lies in the load of the central server~\cite{li2020federated}. The exponential growth in scale and complexity of FL deployments can lead to a critical bottleneck or even single-point failure~\cite{beltran2024fedstellar} at central servers, jeopardizing efficiency and scalability~\cite{kairouz2021advances}.
Another frontier in collaborative learning is Decentralized Federated Learning (DFL)~\cite{NIPS2017_decentralizedLearning}, whose training paradigm shifts from centralized aggregation to decentralized, where each participant acts as both learner and aggregator. This enables scaling to millions of devices without central bottlenecks~\cite{ma2022federated, feng2021blockchain}. However, even worse than CFL's shortcomings, DFL faces communication efficiency challenges, particularly exchange across participants with large-size models (starting from a million in parameter counts). Broadcasting these model updates, then becomes inefficient in massive networks, leading to increased latency and stagnant communication.

Indeed, communication in decentralized environments can result in transmission collisions, network congestion, and sub-optimal information flow without proper scheduling of model updates. Increased concurrent communications negatively affect bandwidth and throughput, potentially saturating the network's data transmission capacity, and causing data packet loss~\cite{jouhari2023survey}. This necessitates retransmission, worsening congestion, and leading to queuing delays at network devices, significantly slowing down communication across silos~\cite{beltran2023decentralized}. 
%Notably, resource-constrained nodes will also make it increasingly strenuous for communication~\cite{imteaj2021survey, chen2021distributed}.
Additionally, transitioning to DFL introduces unforeseen communication challenges, as silos must directly exchange information with all other silos in the network. This is particularly problematic in bandwidth-constrained environments and with large transmission package sizes~\cite{saha2021decentralized}.

To address these issues, Taheri et al. \cite{taheri2020quantized} introduced quantized push-sum, which quantizes large models with fewer bits and passes them over directed graphs, whereas Hashemi et al. \cite{hashemi2021benefits} presented Deli-CoCo, an iterative DFL algorithm that accelerates convergence rates through multiple gossip steps per iteration. Besides, Tang et al. \cite{tang2022gossipfl} proposed GossipFL, enabling each client to communicate with only one peer using a highly sparsified model. Takezawa et al. also successfully proved the effectiveness of their novel topology for fast communication and finite convergence time on multiple datasets.
However, these approaches often either overlook diverse network topologies or fail to effectively manage large package sizes, and require the participation of all clients in every communication iteration. Therefore, the impact of model sizes on communication or latency performance has not been carefully considered in these works.

Recognizing these barriers, this study seeks for the ratio or the trade-off between model size and the performance latency. 
This consideration is formalized via a research question: ``\textit{How does the size of models affect communication latency and efficiency, and how can these factors be optimized to enhance overall DFL system performance?}'' Answering this consideration could have profound implications for DFL. 
%Via experimental results, the size of models significantly impacts communication latency and efficiency in DFL systems, and optimizing these factors is crucial for enhancing overall performance. 
This understanding can also enhance the scalability of DFL frameworks, enabling them to handle large models without performance degradation. Improved communication efficiency translates directly into reduced network bandwidth requirements and computational costs, offering substantial benefits in cross-silo environments where these resources can be costly. Furthermore, these insights can make DFL systems more robust to network variability and disturbances, which is essential given the diverse infrastructure of cross-silo setups~\cite{beltran2023decentralized}. This leads to a more inclusive and robust learning framework. 

Based on our understanding of conventional flooding broadcast, we propose a gossiping approach prioritizing network optimization and congestion reduction over finding the shortest transmission paths within decentralized settings. Leveraging a combination of the Minimum Spanning Tree (MST) and graph coloring algorithm, our primary goal is to reconstruct network structures and schedule communication work for participating nodes, thus ensuring swift and congestion-free communication in varying network topologies and model complexity. 
%we provide an optimal solution for communication 
Our evaluation encompasses four diverse network topologies representing real-world networks, spanning from random to highly structured systems. 
We assess the efficiency of our approach by rigorous experimentation and comparisons with conventional broadcasting, highlighting the potential of our approach to provide uncongested communication within a decentralized learning environment. 

Notably, this study draws on prior research evaluating training performance across various aggregation approaches in heterogeneous networks~\cite{beltran2024fedstellar, nguyen2024wait, maejima2024tram}, which found comparable accuracy maintenance of DFL to CFL in both broadcast and gossip modes. 
Therefore, this work concentrates on assessing the communication efficiency of our proposed method in transmitting different models, particularly seven different variants from EfficcientNet and MobileNet.
It is important to note that all our experiments are run on an ad-hoc network infrastructure setup, using physical routers and devices. 
The evaluation results demonstrated that our proposed method achieves approximately an 8-fold reduction in bandwidth utilization and a 4.4-fold reduction in transfer time when compared to conventional broadcasting methods, with particular emphasis on its efficacy in handling larger models.

The remainder of this work is organized as follows:
Section~\ref{sec:related} discusses literature regarding this work.
The proposed work is presented in Section~\ref{sec:methodologies}.
Section~\ref{sec:experiments} and Section~\ref{sec:eval} provide a detailed description and evaluation of the experiments, respectively.
Finally, Section~\ref{sec:conclusion} concludes this work.

\section{Related Work} \label{sec:related}

Communication between silos in conventional FL can only be done via a central aggregator, which susceptible to be bottlenecks~\cite{beltran2024fedstellar}. However, shifting to DFL raised communication problems when silos have to directly exchange knowledge with all others within the network. This creates significant challenges, particularly in bandwidth-constrained environments and with large transmission package sizes~\cite{saha2021decentralized}. Regarding this, SCAMP~\cite{ganesh2003peer}, a Peer-to-Peer membership protocol for gossip-based systems, dynamically adjusts partial view sizes using logarithmic group size estimates, ensuring reliability and balanced views without global size information. More recently, Liu et al.~\cite{liu2022decentralized} proposes a framework that alternates between multiple local updates and inter-node communications, enhanced by a compressed communication mechanism to improve efficiency while maintaining convergence. However, these approaches often overlook diverse network topologies, fail to account for large package sizes, and require all clients to be active in every communication iteration.

Das et al.~\cite{das2022cross}, on the other hand, emphasized that the system becomes communication-wise expensive when all parties communicate in every training round. They proposed reducing concurrent communication by selecting a subgroup of silos or their gradients for participation in each round, considered an optimal solution. Observing from Machine-Learning-based approaches, most of the works use Deep Reinforcement Learning (e.g., Q-Learning)~\cite{shoab2022intelligent, muthanna2022deep} or gradient sparsification strategies~\cite{wang2023sparsfa}. Although optimized based on adaptively learning from the system environment, these methods face limitations in terms of high computational demands and reliance on specific learning network structures. We also believe that the selection of a subgroup~\cite{hu2019decentralized, wang2023sparsfa} (even though it is done based on sorted top importance) for aggregation still results in some information loss. 

%\subsection{Highlighted solutions comparison}
To distinguish our work from other mainstream solutions, we provide more in-depth comparisons with recently highlighted papers \cite{taheri2020quantized, hashemi2021benefits, tang2022gossipfl, takezawa2024beyond}. 
Taheri et al. \cite{taheri2020quantized} introduced quantized push-sum, which involves quantizing large models with fewer bits and passing them over directed graphs using the push-sum protocol. 
However, while their approach reduces communication overhead without compromising convergence rates, the use of quantization may necessitate additional iterations to attain the same accuracy level as non-quantized methods. It is also noteworthy that despite proposing their method for large model scenarios, the evaluation was conducted on simulated setups with only 10 hidden neurons in the neural networks.
Later, Hashemi et al. \cite{hashemi2021benefits} presented Deli-CoCo, an iterative DFL algorithm using arbitrary communication compression to accelerate convergence rates through multiple gossip steps per iteration. While the authors showcased the effectiveness of their approach through meticulous experiment setups with promising results,  we observed some limitations in their evaluation. Specifically, despite discussing the relevance of large models in the problem definition, their evaluation was limited to 2-layer neural networks with 500 neurons in each layer. This experimented model is considered impractical since contemporary resource-constrained devices (e.g., mobiles) are typically required to train models with more complex yet still lightweight like YOLO, MobileNet, or EfficientNet. Another drawback of this paper is the simulation setup on a single computer with a GPU. In this way, the authors fail to capture the complexities of a real decentralized network and unintendedly undermine the practicality of their results. 
In 2022, Tang et al.~\cite{tang2022gossipfl} successfully proposed GossipFL as a novel sparsification algorithm to enable each client to communicate with only one peer using a highly sparsified model. While their large-scale experiments, involving up to 100 clients, demonstrated promising results, the experiments were limited to relatively simple models, with the largest being ResNet-20 (0.27 million trainable parameters). This fails to capture the real effectiveness of the algorithm when applied to more complex, and practical models commonly used in real-world scenarios.
Most recently, a group of Japanese researchers in \cite{takezawa2024beyond} introduced a novel topology having a fast consensus rate and finite convergence time in DFL. 
Regarding the evaluation, Takezawa et al. successfully proved the effectiveness of their work on multiple datasets and other existing topologies. However, we are again concerned that the communication time observed in their simulated experiments may not accurately reflect in practice with the lack of decentralized deployment information. 
This is where our work comes into play, bridging the gap by providing a deployment setup with real physical devices, and subnet divisions to yield results that closely mirror real-world conditions.

%%%%%%%%%%%%%%%%%%%%%%%%% METHOD %%%%%%%%%%%%%%%%%%%%%%%%%%%%%%%%%%%%%%

\section{Proposed method} 
\label{sec:methodologies}

This section outlines our method, systematically organized into four main subsections. We provide a detailed breakdown of the MOSGU workflow, where M -- Manage connectivity~(Subsection \ref{sec:binarr}), O -- Optimize connectivity~(Subsection \ref{sec:OptimalConnectivity}), S -- Schedule communication~(Subsection \ref{sec:smoothcomm}), and GU -- Gossip and Update recipient's queue~(Subsection \ref{sec:smoothcomm}).

\subsection{M -- Manage connectivity}
\label{sec:binarr}
Before initiating gossip protocols, obtaining the color (representing the active timeslot) of each node is crucial. To achieve this, we first need to get the comprehensive connection information of the entire network with $N$ nodes, presented in the form of a graph structure.

\begin{figure}[h!]
\captionsetup{justification=centering}
   \centering
       \includegraphics[width=.43\textwidth]{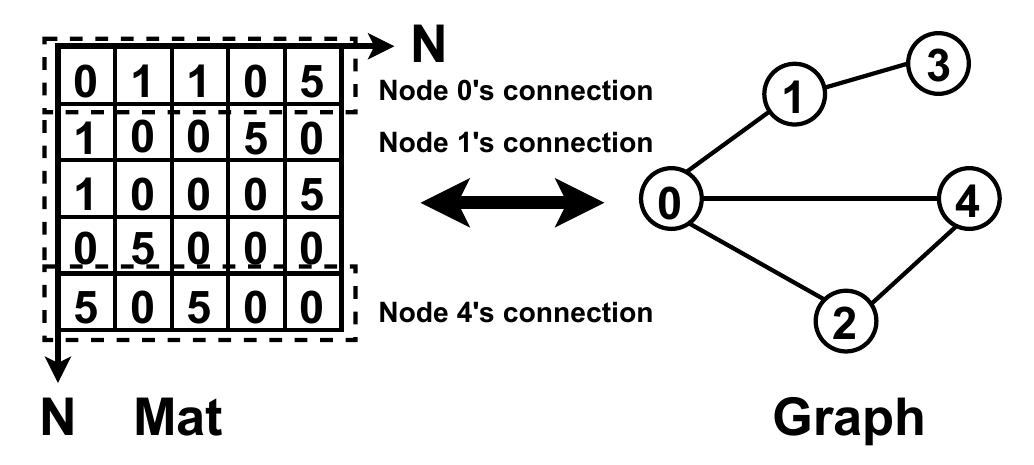} 
   \caption{Adjacency matrix $Mat$ contains communication cost between nodes and its corresponding graph
   }
   \label{fig:mtx2graph}
\end{figure}

Our method employs a dedicated participant within the network, called the moderator, who possesses full connection information and receives real-time system updates. Initially, a random node is selected to serve as the moderator and this node will send a broadcast to inform other participants about its role. Afterward, all nodes transmit their connection information to the moderator, including their IP address and communication cost (calculated based on their actual geographical distance, ping latency, or number of hops) to each connected node. It is noted that although the cost values may slightly differ between two nodes, the moderator will calculate the final cost as the average of those two values in such cases. The moderator then conducts essential graph-related computations, including creating the adjacency matrix $Mat$ (see Fig. \ref{fig:mtx2graph}) based on the received connections' information, constructing an MST $T$ (referred to Section \ref{sec:OptimalConnectivity}), applying a selected coloring algorithm on $T$ and calculating the communication slot length (discussed in Section \ref{sec:smoothcomm}), and finally broadcasting the final color result and neighbor table to each node. 

To preserve decentralization and avoid storing information at any single node, the role of the moderator is periodically rotated among participants in each round of the learning process. Upon delegation of a new moderator, the table with full connection information is forwarded from the old moderator to the new selected one. From the second round onward, the moderator only needs to recompute all graph-related computations and send information to affected nodes when there are changes in the network, such as nodes joining or leaving. If there is no dynamic change, the moderator solely serves as the node keeping the connection information until the next handover to the new periodic moderator. 
This voting matter involves all nodes in the system. Each node casts its vote for the next moderator using the same process as for sending information. Again, the current moderator then aggregates these votes and broadcasts the final result back to all nodes.

This distribution of responsibility helps mitigate the risk of any single node becoming a vulnerability target (e.g., a single-point failure). Besides, while specific methods for determining the moderator in each round are beyond the scope of this study, potential approaches are based on a reputation system, as detailed in \cite{TANG2024269}. These systems assign reputation scores based on the quality of the learning models contributed by participants, ensuring that the node with the highest dedication is trusted to handle sensitive computations.

\subsection{O -- Optimize connectivity}
\label{sec:OptimalConnectivity}
% mst
After creating the adjacency matrix, the moderator constructs the MST to eliminate unnecessary edges or connections while ensuring efficient connectivity between all nodes. 
We consider the complexities (in big $O$ notation) of the three well-known MST algorithms, namely, Kruskal's~\cite{kruskal1956shortest}, Prim's~\cite{prim1957shortest}, and Boruvka's~\cite{boruvka1926jistem}, are as follows: Kruskal's algorithm has a time complexity of $O(E \log E)$, Prim's $O(E + V \log V)$, and Boruvka's $O(E \log V)$. Accordingly, Kruskal's algorithm is generally low complexity for sparse graphs where $E << V^2$, while both Prim's and Boruvka's can efficiently be applied on dense graphs. However, due to its straightforward implementation as well as the advantages of dealing with a high number of nodes in a complete graph, we choose Prim's algorithm. Fig.~\ref{fig:graph-method-mst} illustrates the constructed MST using Prim's algorithm for the input graph G shown in Fig.~\ref{fig:graph-method-orig}.
In regard to scalability, considering MST before coloring can help reduce the computational cost, since finding the optimal coloring can become time-consuming. 
% For example, we can consider using hop counts as the distance metric, with hops counted as the number of nodes, a transmission needs to pass to reach the destination. 

\subsection{S -- Schedule communication}
\label{sec:smoothcomm}
% coloring

After constructing the MST to optimize network structure, the next step in our approach is to schedule communications efficiently. This scheduling is achieved through a coloring algorithm applied to the nodes on the MST. The rationale behind this approach is to minimize bandwidth contention and network congestion by limiting concurrent communication.

\textbf{Algorithm selection:} Among several graph coloring algorithms (Breadth First Search (BFS), DSatur, Welsh-Powell, and Largest Degree First (LDF)), we chose \textit{BFS} for a couple of benefits. 

Firstly, in terms of complexity, \textit{BFS} has the most optimal time complexity $O(V+E)$, which can complete coloring a graph in a very short time. This complexity is considered asymptotically efficient, meaning that its growth rate is significantly slower than $O((V + E) \log V)$ of DSatur $O(VE)$ of LDF and $O(V\log V + E)$ of Welsh-Powell.

Secondly, when comparing coloring results, DSatur may lead to a more balanced distribution and fewer colors used on a standard graph. However, it tends to be more intricate, potentially incurring higher computational costs due to the maintenance of saturation degrees and the utilization of priority queues. Conversely, when coloring an MST, regardless of the algorithm used, the result consistently comprises only two colors. As a consequence, \textit{BFS} becomes the simplest approach for tree coloring. This method entails initiating coloring from a node (randomly chosen as the root) with one color and then assigning the other color to its children, and so forth.

Finally, considering the ease of implementation, \textit{BFS} presents a relatively straightforward logical flow when it only visits each vertex and edge once. Hence, it was chosen as our preferred choice for coloring MSTs, offering advantages in both performance and complexity.

\begin{figure*}[h!]
%\captionsetup{justification=centering}
   \begin{subfigure}{0.3\textwidth}
   \centering
       \includegraphics[scale=0.15]{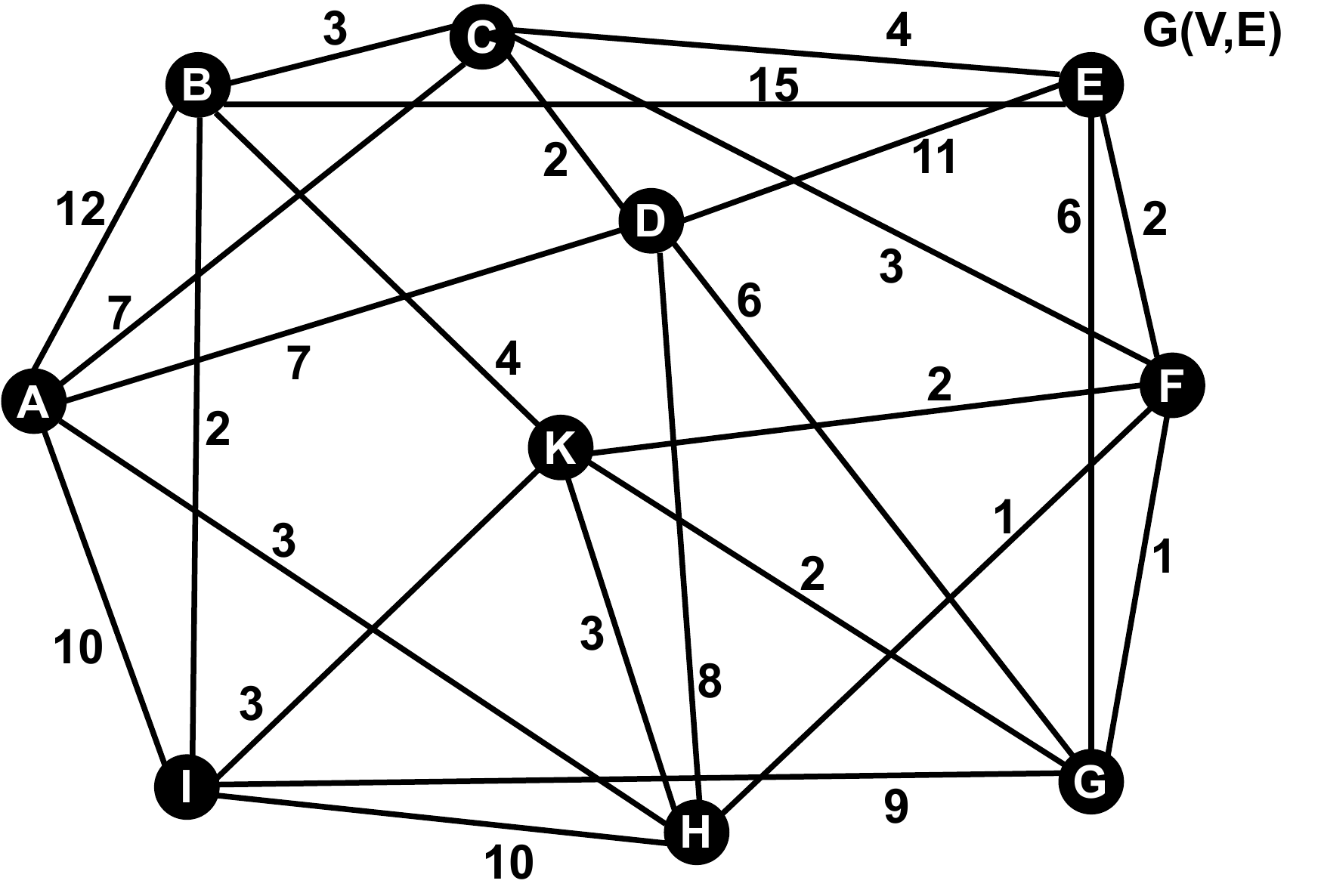} 
       \caption{G is a connected and weighted graph}
       \label{fig:graph-method-orig}
   \end{subfigure}\hfill
   \begin{subfigure}{0.3\textwidth}
   \centering
       \includegraphics[scale=0.15]{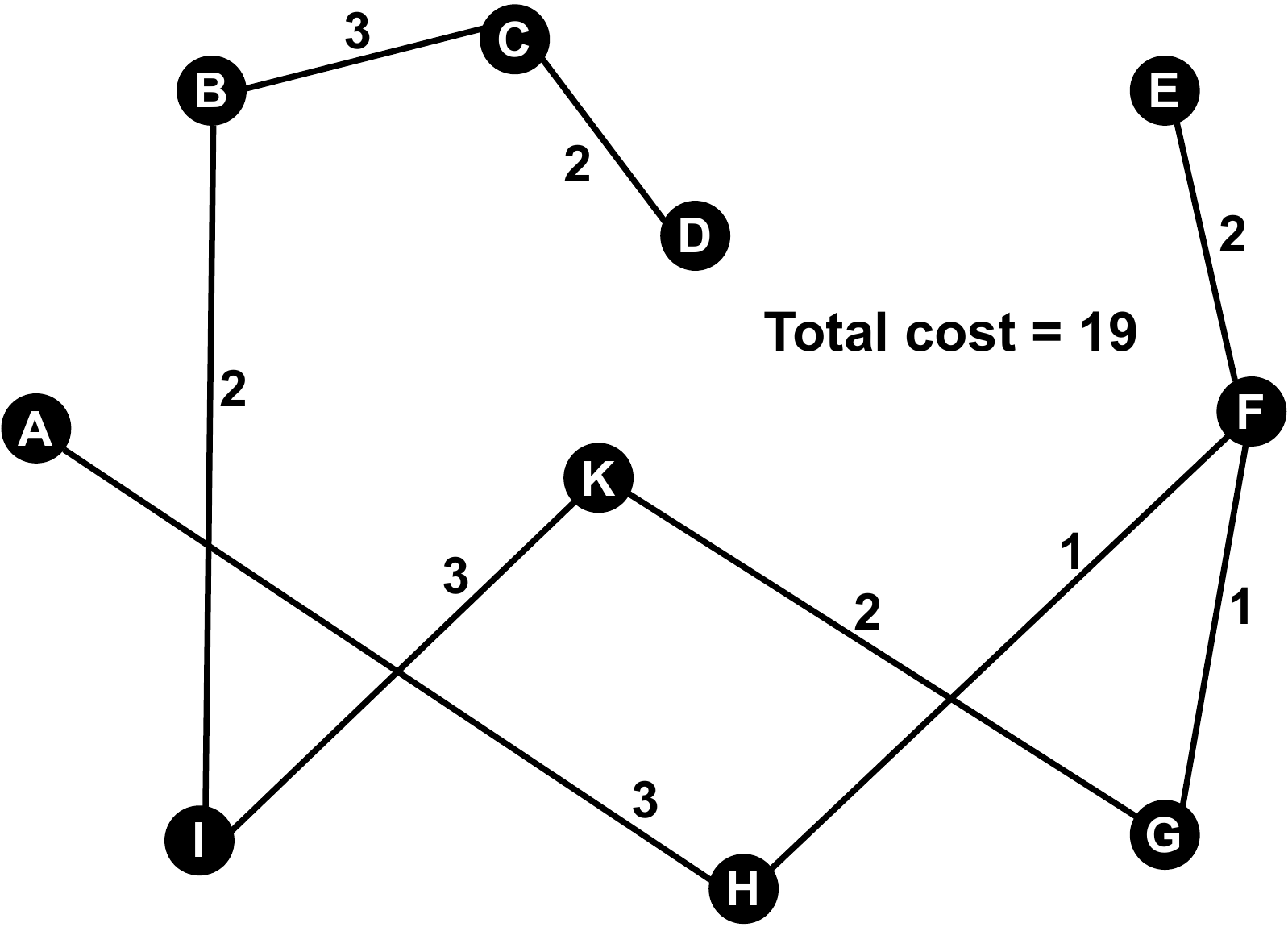}
       \caption{MST on graph G}
       \label{fig:graph-method-mst}
   \end{subfigure}\hfill
   \begin{subfigure}{0.3\textwidth}
   \centering
       \includegraphics[scale=0.15]{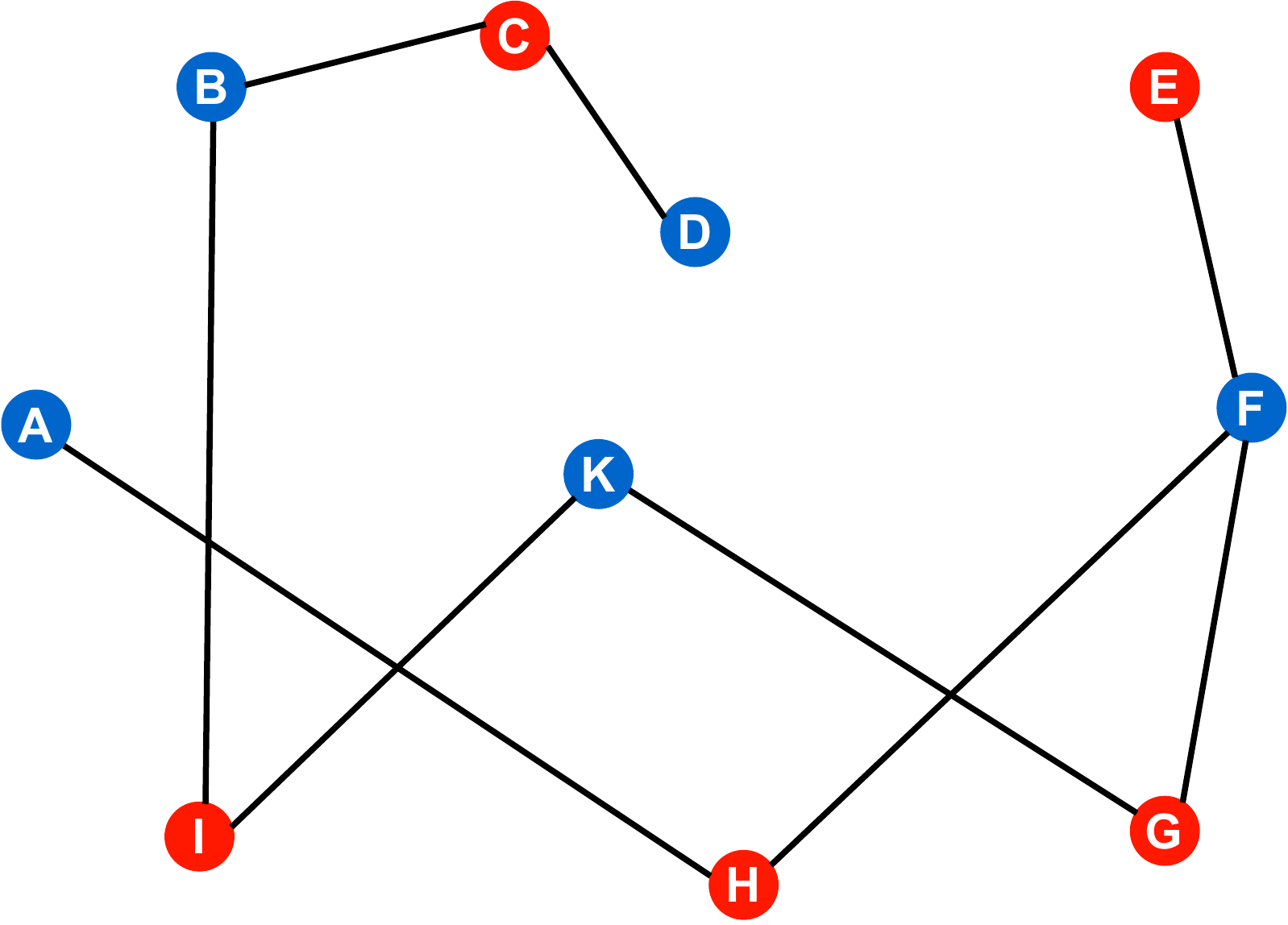}
       \caption{Coloring on  MST}
       \label{fig:graph-method-coloring}
   \end{subfigure}
   \caption{Redundancy removal and coloring vertices on an undirected graph}
   \label{fig:graph-method}
\end{figure*}

Based on the coloring outcome, nodes sharing the same color will operate or transmit simultaneously within the same time slot. Consequently, since red and blue nodes (in Fig. \ref{fig:graph-method-coloring}) cannot function concurrently, this coloring reduces the occurrence of overlapping communications. As a result, it helps in managing the network bandwidth more efficiently by minimizing simultaneous data transmissions.

\textbf{Calculating slot length:} In this work, the time slot length for one color is fixed and calculated based on the ping latency and transferred data's size. Based on the initial reported information, the moderator identifies the max ping value (normally in \textit{ms}) of each node to its neighbors and later finds the highest of these maximum values between nodes having the same color, referred to as $ping_{max}$. Using $ping_{max}$ and the size of the actual model being transmitted $M_{size}$(in megabytes), along with the size of the initial ping data $ping_{size}$(in bytes), we calculate the final length (in second) for each color slot, mathematically formulated as: $slot = \frac{ping_{max} \times M_{size} \times 1000}{ping_{size}} (s)$.

\subsection{GU -- Gossip and Update recipient's queue}
\label{sec:gossipqueue}

Each node contains a FIFO (First In -- First Out) queue $\mathcal{F}$. The queue contains 3-tuples $(O, t, M)$, with $O$ as the identifier of the model's owner (e.g., IP address), $t$ is the training round index, and $M$ indicates model parameters.

Considering that each model $M$ may encompass a vast array of parameters, we have adopted a FIFO ordering system to efficiently manage data flow. Within this system, data entries are processed and dispatched to subsequent recipients based on their sequential order in $\mathcal{F}$. This ensures that models entered earlier are forwarded prior to those that arrive later, maintaining an orderly transmission process. Crucially, to prevent issues of duplicate transmissions, once a model is transmitted or forwarded, it is immediately removed from $\mathcal{F}$. However, if the network temporarily disrupts during transmission, the model will be kept in $\mathcal{F}$ and retransmitted in the node's next turn after it reconnects to the system.

Observe that if the degree of a node in the MST is 1, it connects to only one other node, and receives all entries on its $\mathcal{F}$ from that node. Consequently, apart from locally generated models, a node with degree 1 will not have to forward any received models back to its only neighbor, and its $\mathcal{F}$ queue is always promptly set to be empty after it sends its own model. 

Based on the coloring results (Fig. \ref{fig:graph-method-coloring}), the update on $\mathcal{F}$ on each node, which takes place during model sharing within one communication round of decentralized learning, is presented in \Cref{tab:gossip-example}. 

\begin{table*}[h]
\centering
\caption{$\mathcal{F}$ updates during gossiping process in one round (orange is received models, black is to-be transmitted models)}
\def\arraystretch{1.35}
\label{tab:gossip-example}
\resizebox{\textwidth}{!}{
\begin{tabular}{|c|c|c|c|c|c|c|c|c|c|c|}
\hline
\diagbox{\textbf{Color}}{\textbf{Node}} & \textbf{\textcolor{blue}{A}} & \textbf{\textcolor{blue}{B}} & \textbf{\textcolor{red}{C}} & \textbf{\textcolor{blue}{D}} & \textbf{\textcolor{red}{E}} & \textbf{\textcolor{blue}{F}} & \textbf{\textcolor{red}{G}} & \textbf{\textcolor{red}{H}} & \textbf{\textcolor{red}{I}} & \textbf{\textcolor{blue}{K}} \\ \hline
\textbf{\textcolor{red}{Red}}                                    & A\textcolor{orange}{H}                              & BCI                             & \textcolor{orange}{C}                               & D\textcolor{orange}{C}                            & \textcolor{orange}{E}                               & FEGH                            & \textcolor{orange}{G}                               & \textcolor{orange}{H}                               & \textcolor{orange}{I}                               & KGI                             \\ \hline
\textbf{\textcolor{blue}{Blue}}           & \textcolor{orange}{AH}                              & \textcolor{orange}{B}CI                             & \textcolor{orange}{C}BD                             & \textcolor{orange}{DC}                             & \textcolor{orange}{EF}                              & \textcolor{orange}{F}EGH                            & \textcolor{orange}{G}FK                             & \textcolor{orange}{H}AF                             & \textcolor{orange}{I}BK                             & \textcolor{orange}{K}GI                             \\ \hline
\textbf{\textcolor{red}{Red}}           & \textcolor{orange}{AH}                               & \textcolor{orange}{B}CI                            & \textcolor{orange}{CB}D                                &  \textcolor{orange}{DCB}                              & \textcolor{orange}{EF}                                  & \textcolor{orange}{F}EGHA                                 & \textcolor{orange}{GF}K                                & \textcolor{orange}{HA}F                                & \textcolor{orange}{IB}K                                 & \textcolor{orange}{K}GIFB                                 \\ \hline
\textbf{\textcolor{blue}{Blue}}            & \textcolor{orange}{AH}                               & \textcolor{orange}{BC}I                            & \textcolor{orange}{CB}D                                &  \textcolor{orange}{DCB}                              & \textcolor{orange}{EF}                                  & \textcolor{orange}{FE}GHA                                 & \textcolor{orange}{GF}KE                                & \textcolor{orange}{HA}FE                                & \textcolor{orange}{IB}KCG                                 & \textcolor{orange}{KG}IFB                                 \\ \hline
\textbf{\textcolor{red}{Red}}           & \textcolor{orange}{AHF}                               & \textcolor{orange}{BC}IDK                            & \textcolor{orange}{CBD}                                &  \textcolor{orange}{DCB}                              & \textcolor{orange}{EF}                                  & \textcolor{orange}{FE}GHAK                                 & \textcolor{orange}{GFK}E                                & \textcolor{orange}{HAF}E                                & \textcolor{orange}{IBK}CG                                 & \textcolor{orange}{KG}IFB                                 \\ \hline
\textbf{\textcolor{blue}{Blue}}            & \textcolor{orange}{AHF}                               & \textcolor{orange}{BCI}DK                            & \textcolor{orange}{CBD}I                                &  \textcolor{orange}{DCB}                              & \textcolor{orange}{EFG}                                  & \textcolor{orange}{FEG}HAK                                 & \textcolor{orange}{GFK}EI                                & \textcolor{orange}{HAF}EG                                & \textcolor{orange}{IBK}CG                                 & \textcolor{orange}{KGI}FB                                 \\ \hline
\textbf{\textcolor{red}{Red}}           & \textcolor{orange}{AHFE}                               & \textcolor{orange}{BCI}DK                            & \textcolor{orange}{CBDI}                                &  \textcolor{orange}{DCBI}                              & \textcolor{orange}{EFG}                                  & \textcolor{orange}{FEG}HAK                                 & \textcolor{orange}{GFKE}I                                & \textcolor{orange}{HAFE}G                                & \textcolor{orange}{IBKC}G                                 & \textcolor{orange}{KGI}FBEC                                 \\ \hline
\textbf{\textcolor{blue}{Blue}}            & \textcolor{orange}{AHFE}                               & \textcolor{orange}{BCID}K                            & \textcolor{orange}{CBDI}                                &  \textcolor{orange}{DCBI}                              & \textcolor{orange}{EFGH}                                  & \textcolor{orange}{FEGH}AK                                 & \textcolor{orange}{GFKE}IH                                & \textcolor{orange}{HAFE}G                                & \textcolor{orange}{IBKC}GDF                                 & \textcolor{orange}{KGIF}BEC                                 \\ \hline
\textbf{\textcolor{red}{Red}}           & \textcolor{orange}{AHFEG}                               & \textcolor{orange}{BCID}KG                            & \textcolor{orange}{CBDI}                                &  \textcolor{orange}{DCBI}                              & \textcolor{orange}{EFGH}                                  & \textcolor{orange}{FEGH}AKI                                 & \textcolor{orange}{GFKEI}H                                & \textcolor{orange}{HAFEG}                                & \textcolor{orange}{IBKCG}DF                                 & \textcolor{orange}{KGIF}BEC                                 \\ \hline
\textbf{\textcolor{blue}{Blue}}            & \textcolor{orange}{AHFEG}                               & \textcolor{orange}{BCIDK}G                            & \textcolor{orange}{CBDI}K                                &  \textcolor{orange}{DCBI}                              & \textcolor{orange}{EFGHA}                                  & \textcolor{orange}{FEGHA}KI                                 & \textcolor{orange}{GFKEI}HAB                                & \textcolor{orange}{HAFEG}                                & \textcolor{orange}{IBKCG}DF                                 & \textcolor{orange}{KGIFB}EC                                 \\ \hline
\textbf{\textcolor{red}{Red}}           & \textcolor{orange}{AHFEG}                               & \textcolor{orange}{BCIDK}G                            & \textcolor{orange}{CBDIK}                                &  \textcolor{orange}{DCBIK}                              & \textcolor{orange}{EFGHA}                                  & \textcolor{orange}{FEGHA}KI                                 & \textcolor{orange}{GFKEIH}AB                                & \textcolor{orange}{HAFEG}                                & \textcolor{orange}{IBKCGD}F                                 & \textcolor{orange}{KGIFB}ECHD                                 \\ \hline
\textbf{\textcolor{blue}{Blue}}            & \textcolor{orange}{AHFEG}                               & \textcolor{orange}{BCIDKG}                            & \textcolor{orange}{CBDIK}G                                &  \textcolor{orange}{DCBIK}                              & \textcolor{orange}{EFGHAK}                                  & \textcolor{orange}{FEGHAK}I                                 & \textcolor{orange}{GFKEIH}AB                                & \textcolor{orange}{HAFEG}K                                & \textcolor{orange}{IBKCGD}FE                                 & \textcolor{orange}{KGIFBE}CHD                                 \\ \hline
\textbf{\textcolor{red}{Red}}           & \textcolor{orange}{AHFEGK}                               & \textcolor{orange}{BCIDKG}F                            & \textcolor{orange}{CBDIKG}                                &  \textcolor{orange}{DCBIKG}                              & \textcolor{orange}{EFGHAK}                                  & \textcolor{orange}{FEGHAK}I                                 & \textcolor{orange}{GFKEIHA}B                                & \textcolor{orange}{HAFEGK}                                & \textcolor{orange}{IBKCGDF}E                                 & \textcolor{orange}{KGIFBE}CHDA                                 \\ \hline
\textbf{\textcolor{blue}{Blue}}            & \textcolor{orange}{AHFEGK}                               & \textcolor{orange}{BCIDKGF}                            & \textcolor{orange}{CBDIKG}F                                &  \textcolor{orange}{DCBIKG}                              & \textcolor{orange}{EFGHAKI}                                  & \textcolor{orange}{FEGHAKI}                                 & \textcolor{orange}{GFKEIHA}BC                                & \textcolor{orange}{HAFEGK}I                                & \textcolor{orange}{IBKCGDF}E                                 & \textcolor{orange}{KGIFBEC}HDA                                 \\ \hline
\textbf{\textcolor{red}{Red}}           & \textcolor{orange}{AHFEGKI}                               & \textcolor{orange}{BCIDKGF}E                            & \textcolor{orange}{CBDIKGF}                                &  \textcolor{orange}{DCBIKGF}                              & \textcolor{orange}{EFGHAKI}                                  & \textcolor{orange}{FEGHAKI}B                                 & \textcolor{orange}{GFKEIHAB}C                                & \textcolor{orange}{HAFEGKI}                                & \textcolor{orange}{IBKCGDFE}                                 & \textcolor{orange}{KGIFBEC}HDA                                 \\ \hline
\textbf{\textcolor{blue}{Blue}}            & \textcolor{orange}{AHFEGKI}                               & \textcolor{orange}{BCIDKGFE}                            & \textcolor{orange}{CBDIKGF}E                                &  \textcolor{orange}{DCBIKGF}                              & \textcolor{orange}{EFGHAKIB}                                  & \textcolor{orange}{FEGHAKIB}                                 & \textcolor{orange}{GFKEIHAB}C                                & \textcolor{orange}{HAFEGKI}B                                & \textcolor{orange}{IBKCGDFE}H                                 & \textcolor{orange}{KGIFBECH}DA                                 \\ \hline
\textbf{\textcolor{red}{Red}}           & \textcolor{orange}{AHFEGKIB}                               & \textcolor{orange}{BCIDKGFE}H                            & \textcolor{orange}{CBDIKGFE}                                &  \textcolor{orange}{DCBIKGFE}                              & \textcolor{orange}{EFGHAKIB}                                  & \textcolor{orange}{FEGHAKIB}C                                 & \textcolor{orange}{GFKEIHABC}                                & \textcolor{orange}{HAFEGKIB}                                & \textcolor{orange}{IBKCGDFEH}                                 & \textcolor{orange}{KGIFBECH}DA                                 \\ \hline
\textbf{\textcolor{blue}{Blue}}            & \textcolor{orange}{AHFEGKIB}                               & \textcolor{orange}{BCIDKGFEH}                            & \textcolor{orange}{CBDIKGFE}H                                &  \textcolor{orange}{DCBIKGFE}                              & \textcolor{orange}{EFGHAKIBC}                                  & \textcolor{orange}{FEGHAKIBC}                                 & \textcolor{orange}{GFKEIHABC}D                                & \textcolor{orange}{HAFEGKIB}C                                & \textcolor{orange}{IBKCGDFEH}                                 & \textcolor{orange}{KGIFBECHD}A                                 \\ \hline
\textbf{\textcolor{red}{Red}}           & \textcolor{orange}{AHFEGKIBC}                               & \textcolor{orange}{BCIDKGFEH}                            & \textcolor{orange}{CBDIKGFEH}                                &  \textcolor{orange}{DCBIKGFEH}                              & \textcolor{orange}{EFGHAKIBC}                                  & \textcolor{orange}{FEGHAKIBC}D                                 & \textcolor{orange}{GFKEIHABCD}                                & \textcolor{orange}{HAFEGKIBC}                                & \textcolor{orange}{IBKCGDFEH}                                 & \textcolor{orange}{KGIFBECHD}A                                 \\ \hline
\textbf{\textcolor{blue}{Blue}}            & \textcolor{orange}{AHFEGKIBC}                               & \textcolor{orange}{BCIDKGFEH}                            & \textcolor{orange}{CBDIKGFEH}                                &  \textcolor{orange}{DCBIKGFEH}                              & \textcolor{orange}{EFGHAKIBCD}                                  & \textcolor{orange}{FEGHAKIBCD}                                 & \textcolor{orange}{GFKEIHABCD}                                & \textcolor{orange}{HAFEGKIBC}D                                & \textcolor{orange}{IBKCGDFEH}A                                 & \textcolor{orange}{KGIFBECHDA}                                 \\ \hline
\textbf{\textcolor{red}{Red}}           & \textcolor{orange}{AHFEGKIBCD}                               & \textcolor{orange}{BCIDKGFEH}A                            & \textcolor{orange}{CBDIKGFEH}                                &  \textcolor{orange}{DCBIKGFEH}                              & \textcolor{orange}{EFGHAKIBCD}                                  & \textcolor{orange}{FEGHAKIBCD}                                 & \textcolor{orange}{GFKEIHABCD}                                & \textcolor{orange}{HAFEGKIBCD}                                & \textcolor{orange}{IBKCGDFEHA}                                 & \textcolor{orange}{KGIFBECHDA}                                 \\ \hline
\textbf{\textcolor{blue}{Blue}}           & \textcolor{orange}{AHFEGKIBCD}                               & \textcolor{orange}{BCIDKGFEHA}                            & \textcolor{orange}{CBDIKGFEH}A                                &  \textcolor{orange}{DCBIKGFEH}                              & \textcolor{orange}{EFGHAKIBCD}                                  & \textcolor{orange}{FEGHAKIBCD}                                 & \textcolor{orange}{GFKEIHABCD}                                & \textcolor{orange}{HAFEGKIBCD}                                & \textcolor{orange}{IBKCGDFEHA}                                 & \textcolor{orange}{KGIFBECHDA}                                 \\ \hline
\textbf{\textcolor{red}{Red}}           & \textcolor{orange}{AHFEGKIBCD}                               & \textcolor{orange}{BCIDKGFEHA}                            & \textcolor{orange}{CBDIKGFEHA}                                &  \textcolor{orange}{DCBIKGFEHA}                              & \textcolor{orange}{EFGHAKIBCD}                                  & \textcolor{orange}{FEGHAKIBCD}                                 & \textcolor{orange}{GFKEIHABCD}                                & \textcolor{orange}{HAFEGKIBCD}                                & \textcolor{orange}{IBKCGDFEHA}                                 & \textcolor{orange}{KGIFBECHDA}                                 \\ \hline
\end{tabular}
}
\end{table*}

On the first row of the \Cref{tab:gossip-example}, we observe that, as it is the red's turn, all blue nodes (A, B, D, F, K) remain stationary, while the red nodes (C, E, G, H, I) start transmitting the oldest models in their $\mathcal{F}$ to their connections. In this table, orange represents models stored within the system and have already been transmitted to connected nodes, while black denotes models within $\mathcal{F}$, awaiting transmission. The $\mathcal{F}$ is organized from left to right, where ``BCI'' indicates that ``B'' is the oldest model, and ``I'' is the most recent one. Therefore, in the subsequent transmission (during the blue team's turn), model ``B'' will be transmitted. This table continues until all nodes have received all ten models, resulting in all models being denoted in orange. It is important to note that this table illustrates $\mathcal{F}$ activities within a single communication round. Consequently, we simplify the representation by using only the node's name to indicate its local model, omitting the round index.

%%%%%%%%%%%%%%%%%%%%%%%%% EXPERIMENT %%%%%%%%%%%%%%%%%%%%%%%%%%%%%%%%%%%%%%
\section{Experiments}   
\label{sec:experiments}
\begin{figure}[h!]
\captionsetup{justification=centering}
   \centering
       \includegraphics[width=.8\columnwidth]{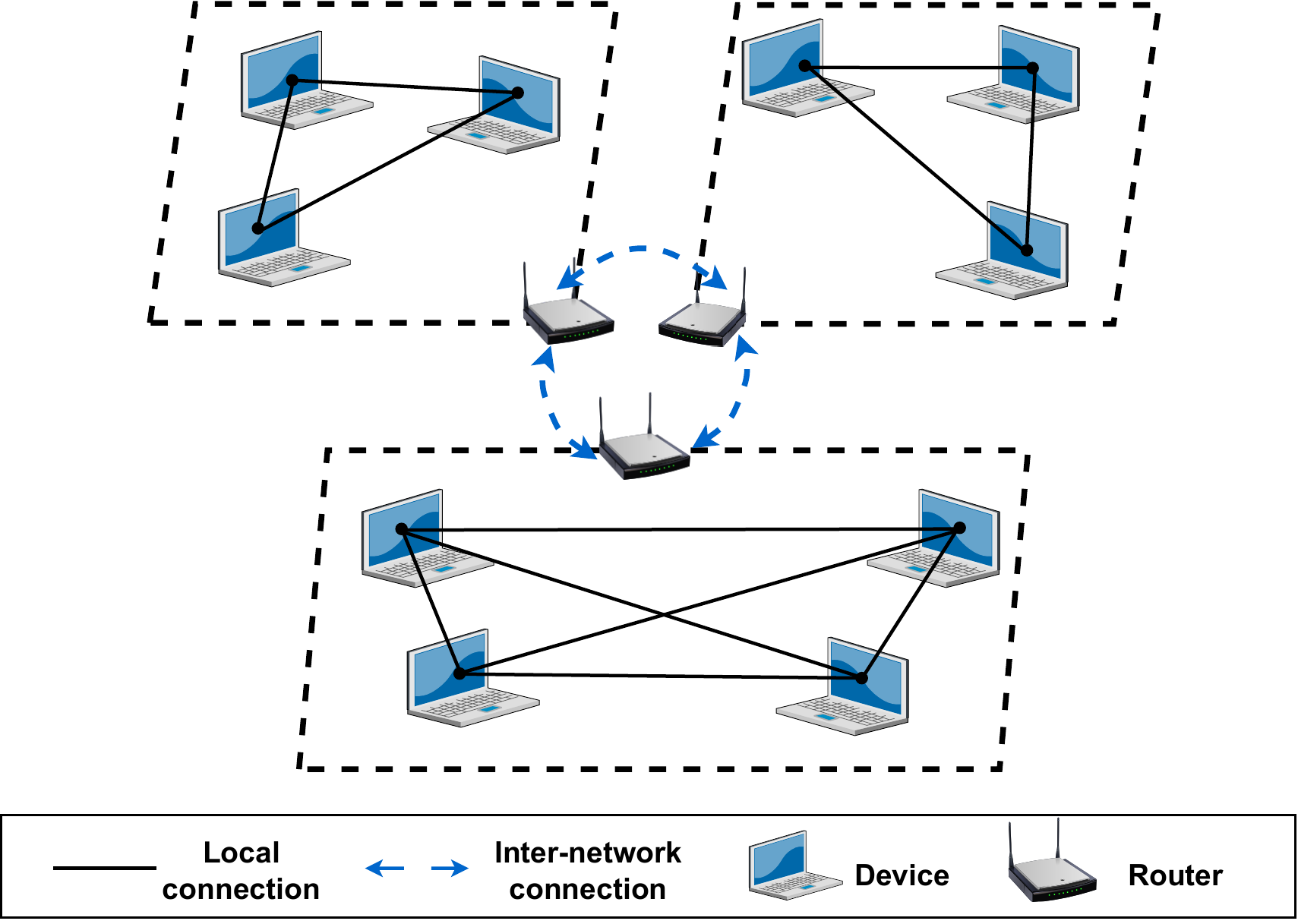} 
   \caption{Experimental setting}
   \label{fig:router-exp}
\end{figure}

\subsection{Hardware and configurations}
We set up ten nodes (N=10), representing ten learning devices handled fully connected by three routers (as shown in Fig. \ref{fig:router-exp}). Each node in this decentralized network has the same low computation power with 4 GB RAM, dual-core processors, and 64 GB of storage, running Ubuntu 22.04. These devices have no GPU. It is noted that these computational configurations align with or are even below the hardware specifications commonly found in contemporary mobile devices. Besides, we utilized File Transfer Protocol (FTP)\footnote{https://www.rfc-editor.org/info/rfc959} for transmitting information and models.

To emulate varying geographical distances and inter-node communication scenarios, we used three routers, each configured to govern a distinct subnetwork. Consequently, data transmission between devices in different subnetworks requires multiple hops within the underlay infrastructure. This routing path comprises initial forwarding from the source device to its designated subnetwork router (source router), followed by relaying through the destination router, and finally to the target device. All three routers in this setup operate at the same transmission speeds. We use ping latency as the distance (cost) between nodes in our experiment.

\begin{figure*}[h]
\captionsetup{justification=centering}
   \begin{subfigure}{0.23\textwidth}
   \centering
       \includegraphics[scale=0.11]{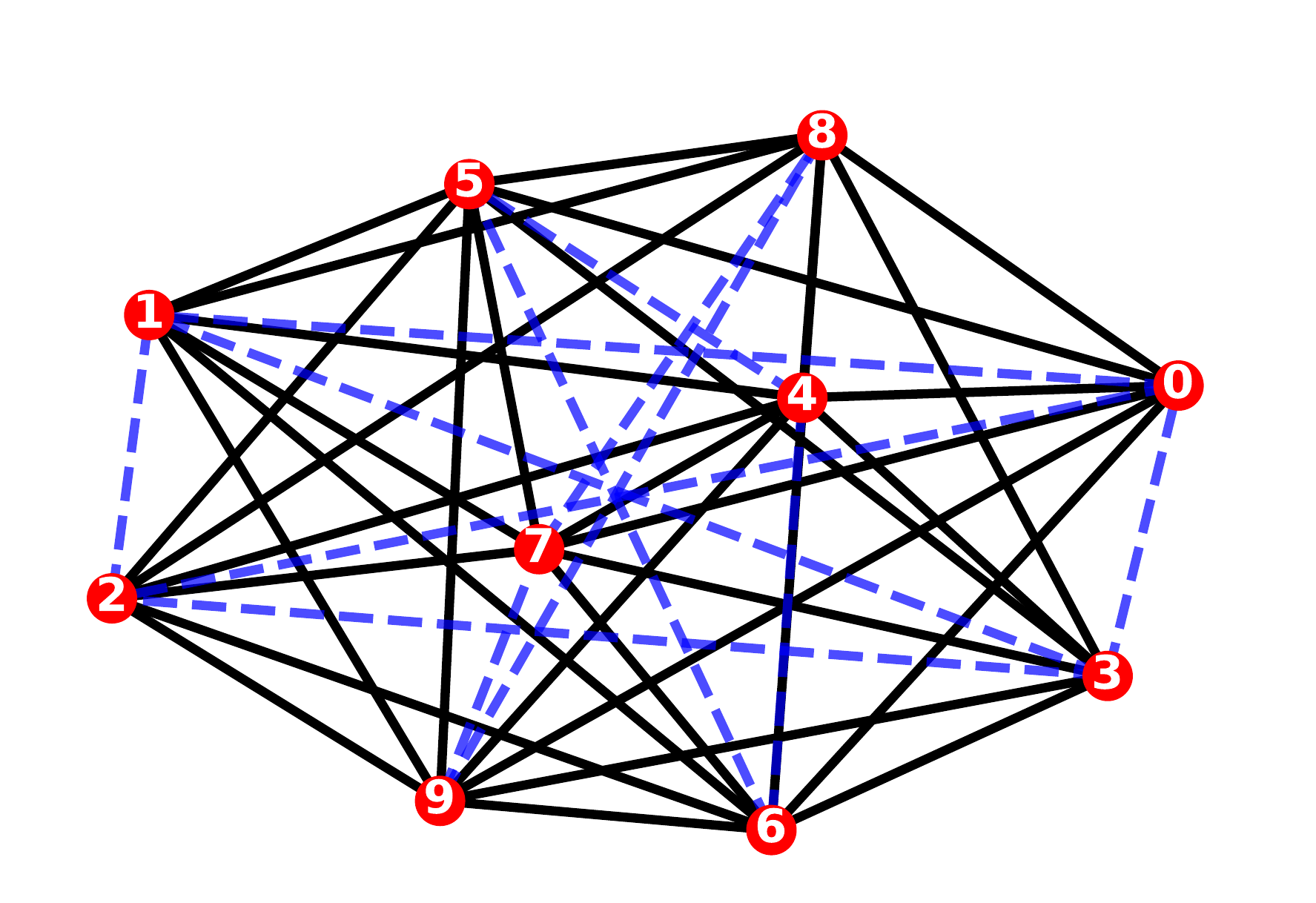} 
       \caption{Complete graph}
       \label{fig:graph-exp-overlay}
   \end{subfigure}\hfill
   \begin{subfigure}{0.24\textwidth}
   \centering
       \includegraphics[scale=0.11]{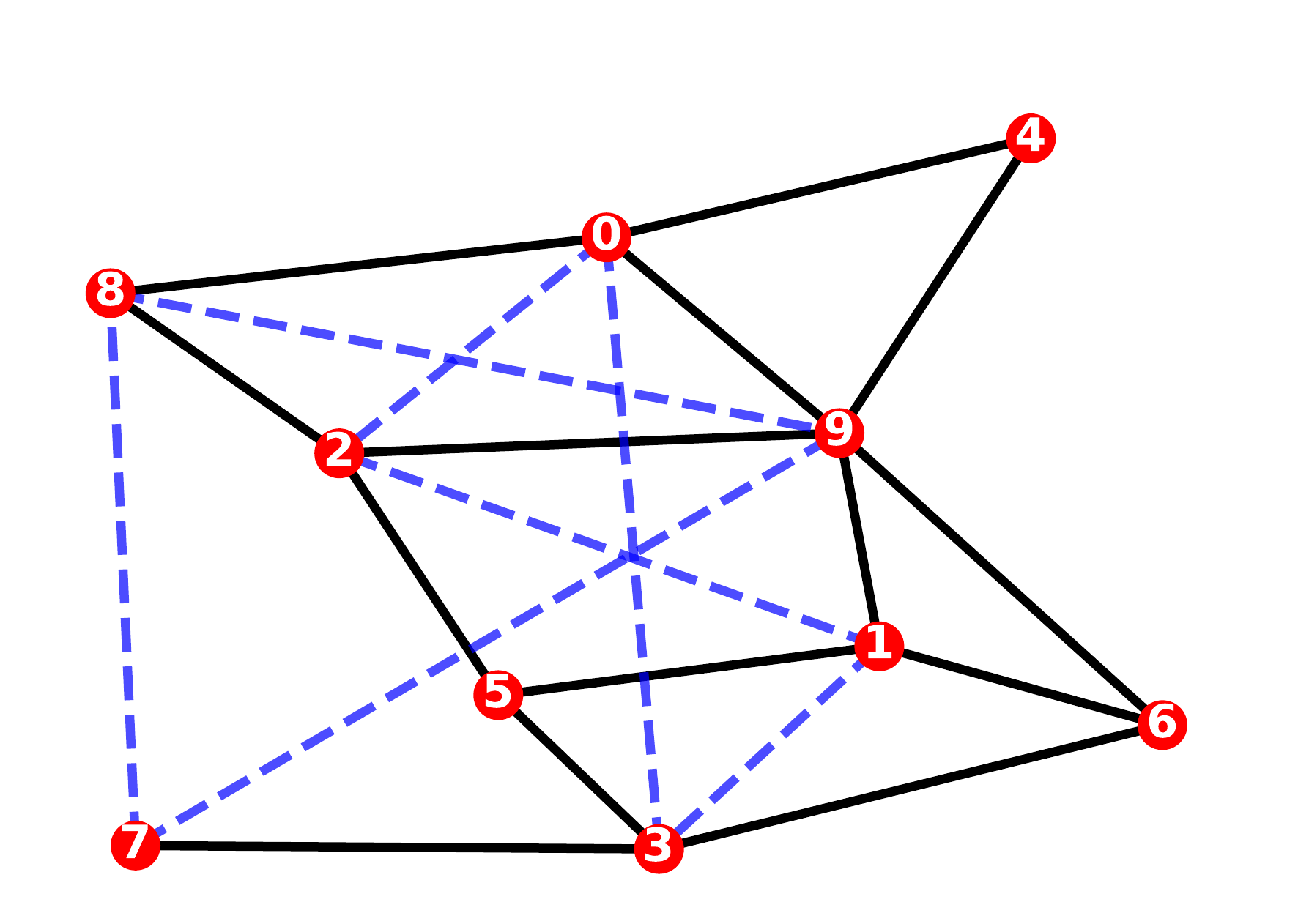}
       \caption{Erdös-Rényi}
       \label{fig:graph-exp-renyi}
   \end{subfigure}\hfill
   \begin{subfigure}{0.25\textwidth}
   \centering
       \includegraphics[scale=0.11]{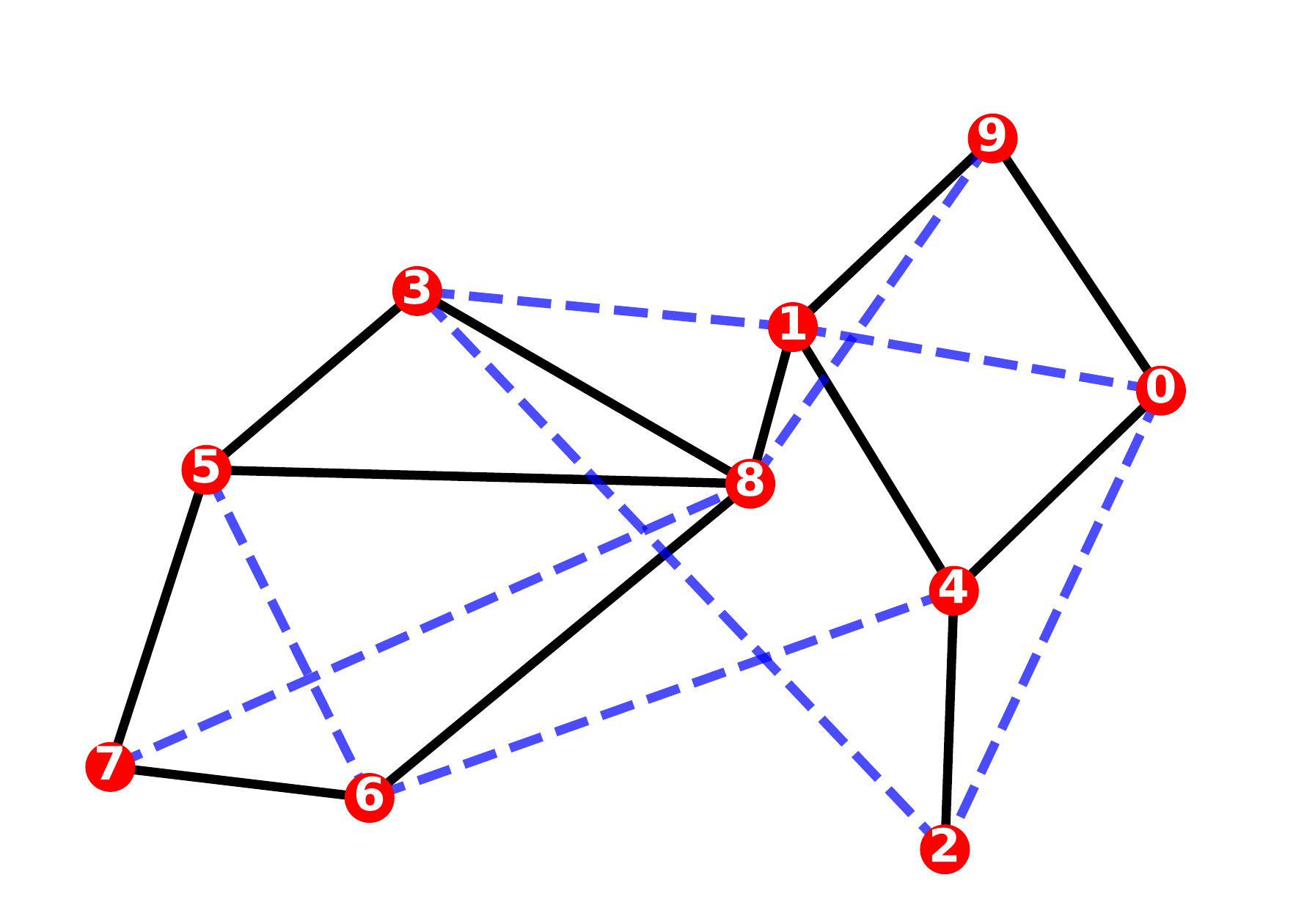}
       \caption{Watts–Strogatz}
       \label{fig:graph-exp-watt}
   \end{subfigure}
    \begin{subfigure}{0.23\textwidth}
   \centering
       \includegraphics[scale=0.11]{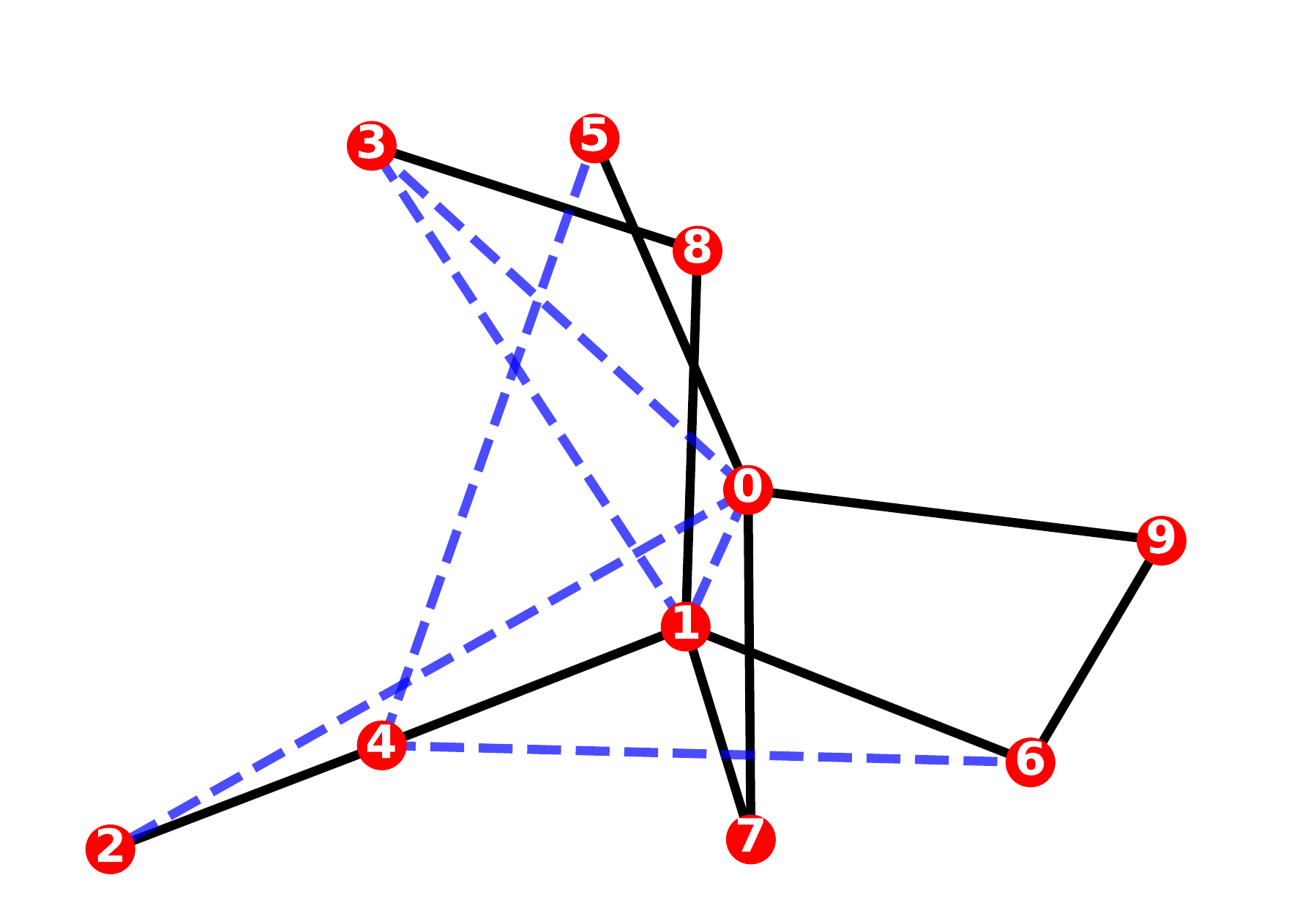}
       \caption{Barabási–Albert}
       \label{fig:graph-exp-bara}
   \end{subfigure}
   \caption{Graph experiments (black edge is interconnection, dashed-blue is local connections)}
   \label{fig:graph-exp}
\end{figure*}
\subsection{Graphs topology}
% bara m=2 = Number of edges to attach from a new node to existing nodes
% watt k=4 = Each node is joined with its `k` nearest neighbors in a ring | p =0.2 = The probability of rewiring each edge
%         topology.
% renyi p=0.5 = Probability for edge creation

Each node in our experiment will connect to every other node. This results in a complete graph on the overlay network. In the underlay network, on the other hand, we experiment with one complete (every node has connections with others) and three random topologies following: Erdös-Rényi \cite{erdos59a}, Watts–Strogatz \cite{watts1998collective}, and Barabási–Albert \cite{barabasi99}. 

The employment of these topologies effectively mirrors the diverse structural characteristics observed in various real-world networks, ranging from random to highly structured systems. 
The Erdös-Rényi model, with its randomly formed connections, is akin to decentralized networks, where links are formed in a random and non-deterministic manner. However, this topology is more of a theoretical baseline than a precise reflection of any specific real-world network due to its simplicity.
The Watts–Strogatz model, capturing the small-world phenomenon, is suitable for networks with high clustering and short path lengths, resembling social networks where individuals in a tightly-knit acquaintance group can access distant parts of the network only by a few steps. 
Lastly, the Barabási–Albert model describes scale-free networks, typical of the Internet's structure, where certain nodes act as highly connected hubs, following a power-law degree distribution and having significantly more connections than others. 
See Fig. \ref{fig:graph-exp} for illustrations of such topologies.

\subsection{Capacity}
In the model selection stage, we consider using multiple model architectures tailored for mobile and low-power devices, such as MobileNet and EfficientNet. Considering the hardware facilities, for EfficientNet, we limit our experimentation to the B0-B3 models, as B4-B7 are considerably heavy and slow without GPU. The model size is measured by the number of parameters in millions or its capacity in Megabytes (MB). These sizes are then categorized into 3 types: small (0-15), and, medium (15.1-30), and large (over 30) based on the MB value. See Table \ref{tab:model-capacity} for model information in detail.

% The 32-bit floating point (FP32) occupies 32 bits (4 bytes) in memory to store one single precision float. The model size, calculated in bytes is then mathematically formalized as: $Size  = n_{params} \times 4$.

\begin{table}[h]
\centering
\caption{Models comparison in terms of number of parameters and capacity}
\label{tab:model-capacity}
\def\arraystretch{1.15}
\resizebox{\columnwidth}{!}{
\begin{tabular}{|l|c|c|c|c|}
\hline
\multicolumn{1}{|c|}{\textbf{Model}} & \multicolumn{1}{c|}{\textbf{Code}} &\textbf{\begin{tabular}[c]{@{}c@{}}\# Params\\ (Millions)\end{tabular}} & \textbf{\begin{tabular}[c]{@{}c@{}}Capacity\\ (MB)\end{tabular}} & \multicolumn{1}{c|}{\textbf{Category}}\\ \hline
EfficientNet-B0                         &b0 & 5.3                                                                     & 21.2 & medium                                                            \\ \hline
EfficientNet-B1                      &b1 & 7.8                                                                     & 31.2  & large                                                           \\ \hline
EfficientNet-B2                      &b2   & 9.2                                                                     & 36.8 & large                                                           \\ \hline
EfficientNet-B3                      &b3 & 12                                                                      & 48 & large                                                              \\ \hline

MobileNetV2                          &v2   & 3.5                                                                     & 14 & small                                                              \\ \hline
MobileNetV3 Small (1.0)              &v3s & 2.9                                                                     & 11.6 & small                                                            \\ \hline
MobileNetV3 Large (1.0)              &v3l & 5.4                                                                     & 21.6 & medium
\\ \hline
\end{tabular}
}
\end{table}

%%%%%%%%%%%%%%%%%%%%%%%%% EVALUATION %%%%%%%%%%%%%%%%%%%%%%%%%%%%%%%%%%%%%%
\section{Evaluation}
\label{sec:eval}
This section analyzes the efficiency evaluation of our proposed method compared to the traditional flooding broadcast mechanism~\cite{lim2001flooding}. 
%\todo[inline]{write a bit more about the broadcast working here}
The performance is assessed through 3 indicators (bandwidth, single transfer time, and total time for a communication round) across different model sizes and network topologies. 
In which, the report on bandwidth (MB/s), details the network's data transmission capabilities, specifically the rate (in megabytes) at which data can be transmitted or received over the network within a specific amount of time (one second). The single transfer time (s) refers to the duration (in seconds) it takes for a local model to be sent from one node to another by using our approach. Finally, the report on the total time for one communication round (s) quantifies the complete duration (in seconds) required for all model updates to be transferred and gossiped across the network.

\begin{figure}[h!]
\captionsetup{justification=centering}
   \begin{subfigure}{0.23\textwidth}
   \centering
       \includegraphics[scale=0.11]{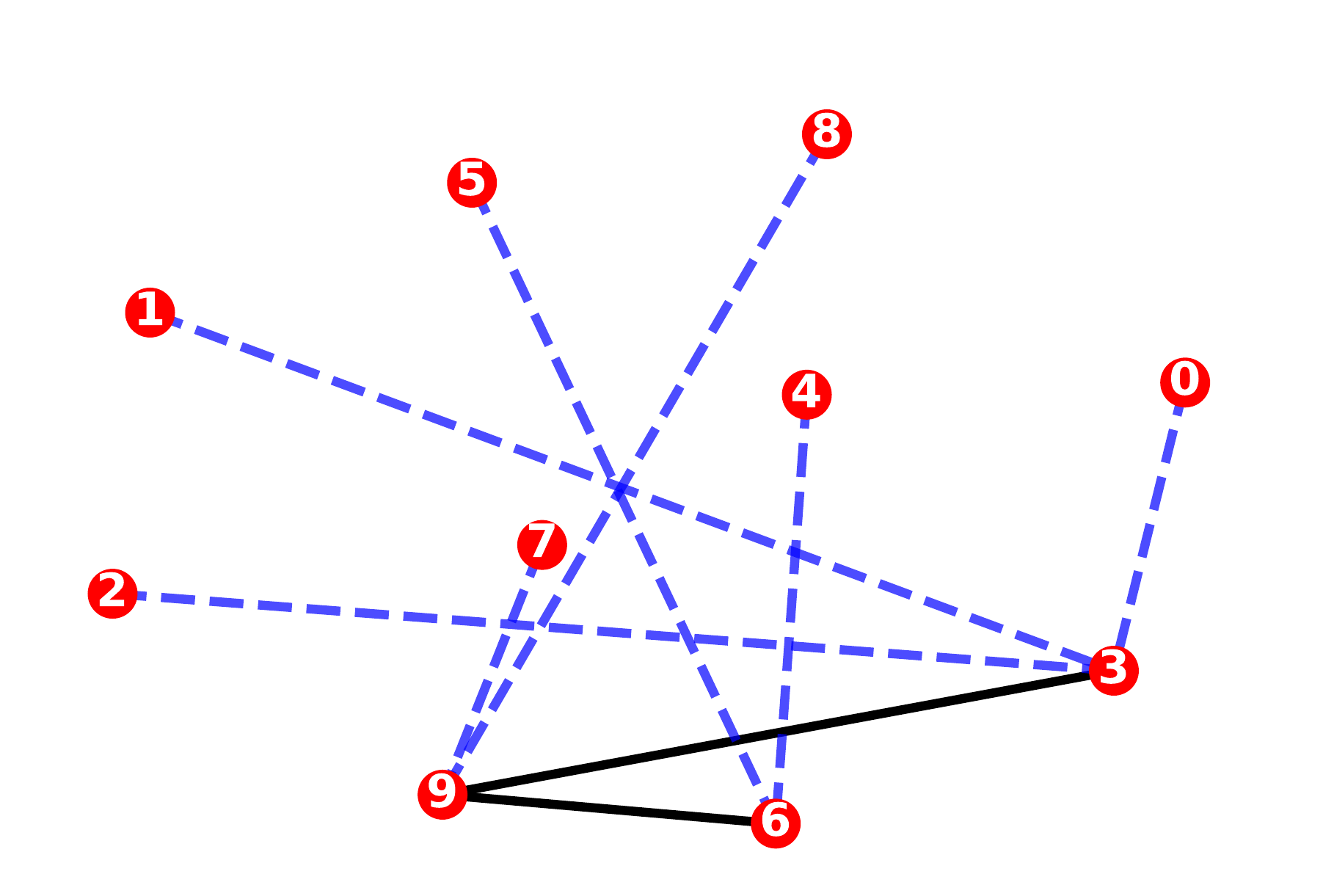} 
       \caption{Complete graph}
       \label{fig:mst-exp-complete}
   \end{subfigure}\hfill
   \begin{subfigure}{0.24\textwidth}
   \centering
       \includegraphics[scale=0.11]{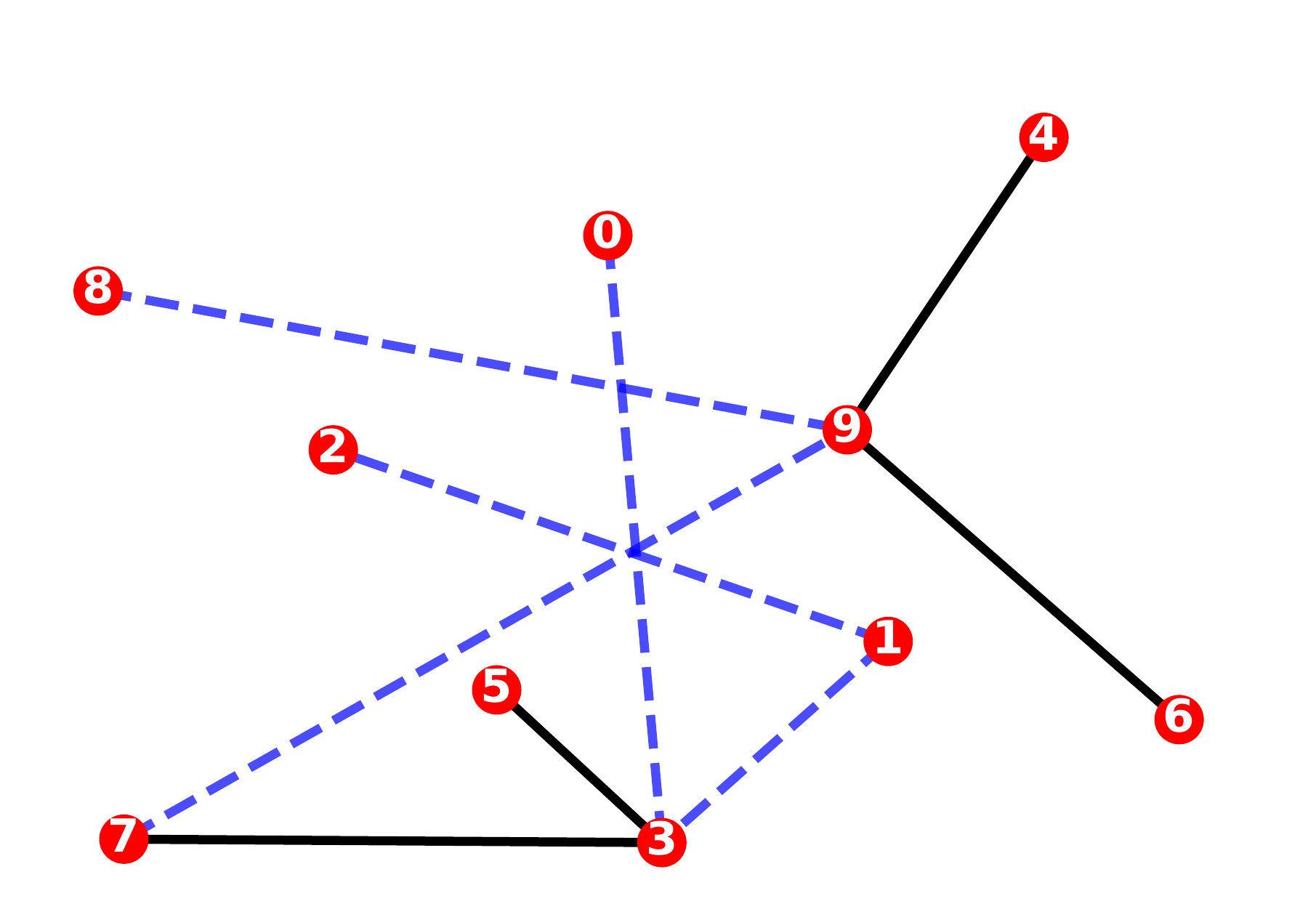}
       \caption{Erdös-Rényi}
       \label{fig:mst-exp-renyi}
   \end{subfigure}\hfill
   \begin{subfigure}{0.25\textwidth}
   \centering
       \includegraphics[scale=0.11]{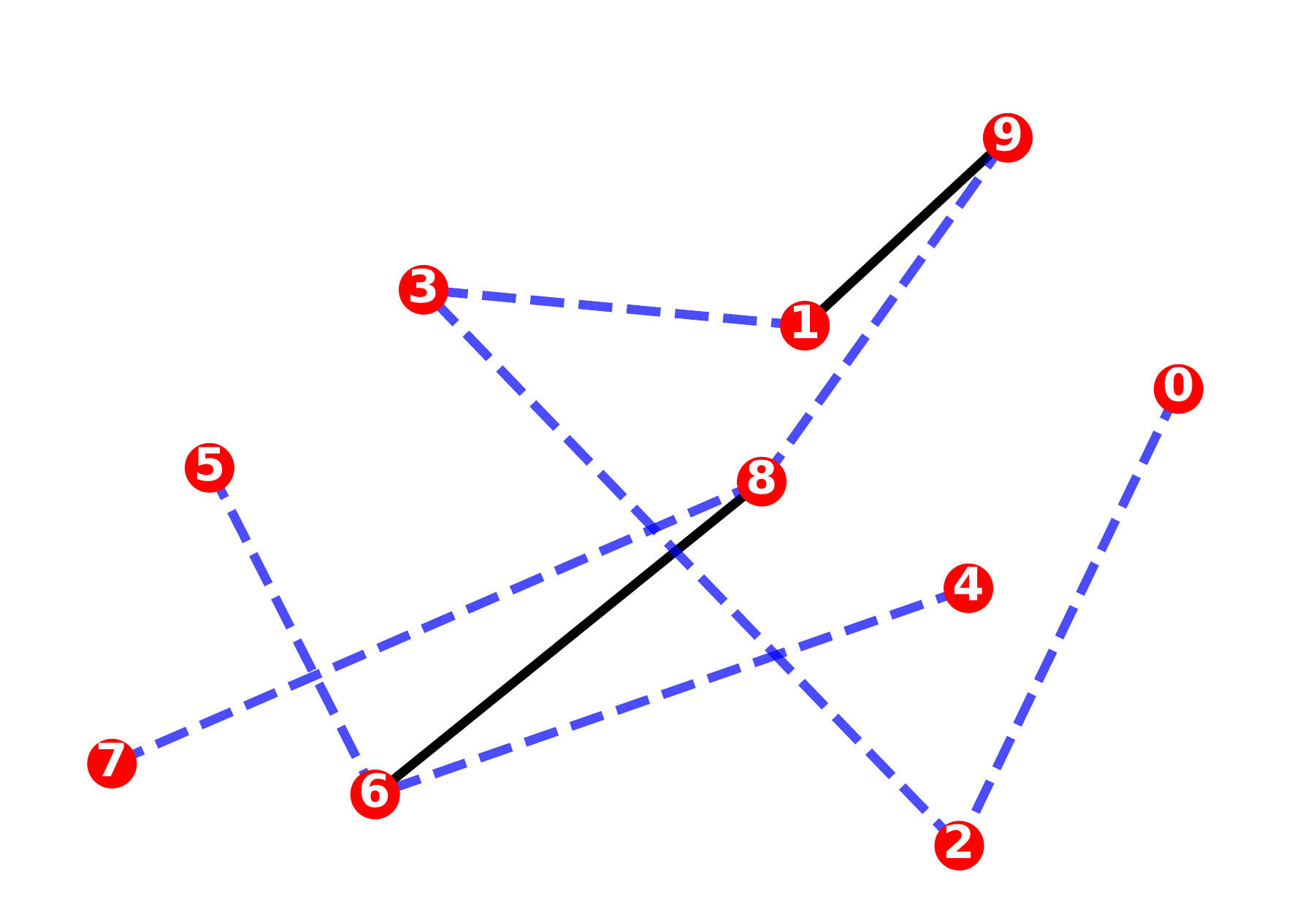}
       \caption{Watts–Strogatz}
       \label{fig:mst-exp-watt}
   \end{subfigure}
    \begin{subfigure}{0.23\textwidth}
   \centering
       \includegraphics[scale=0.11]{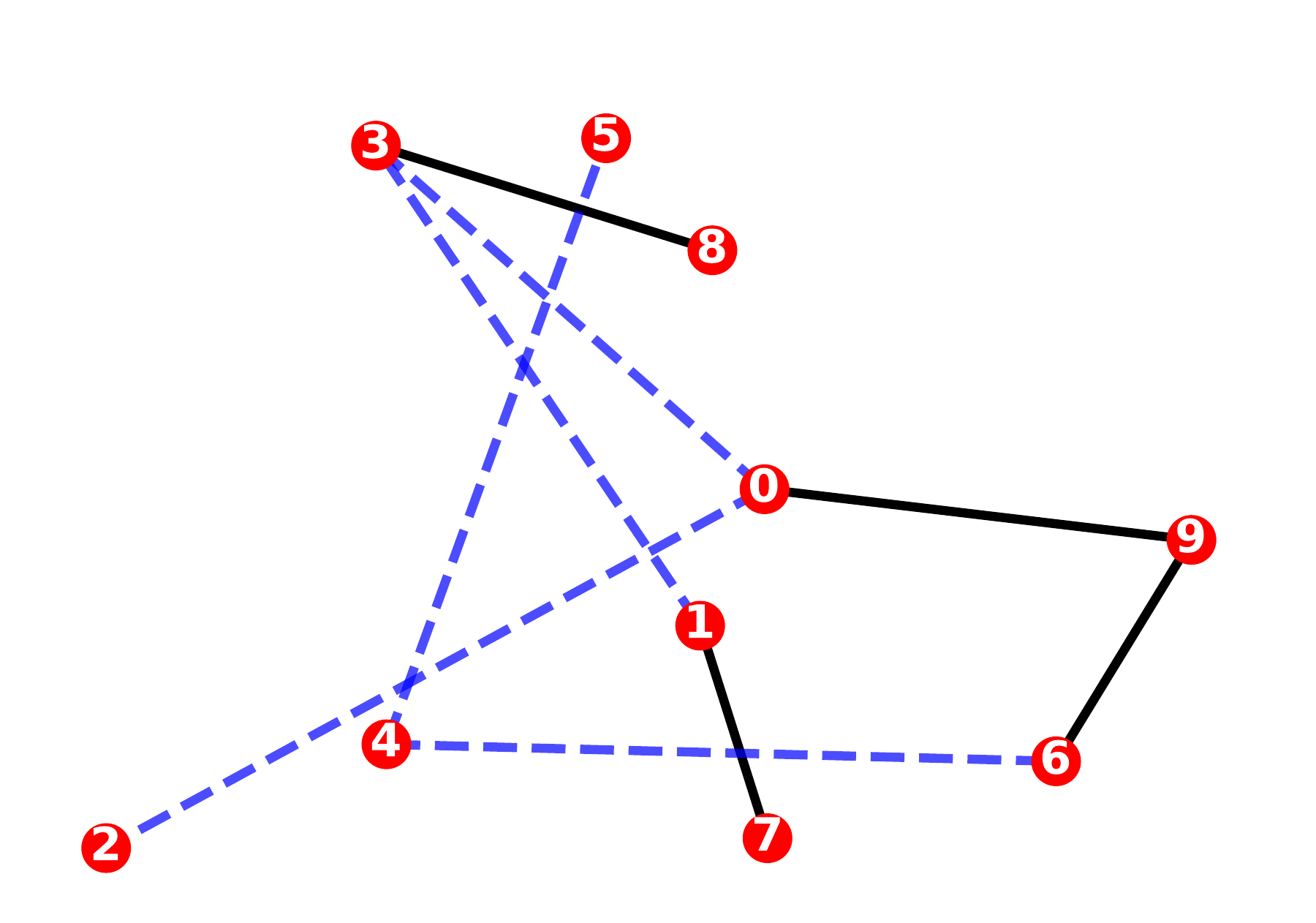}
       \caption{Barabási–Albert}
       \label{fig:mst-exp-bara}
   \end{subfigure}
   \caption{Constructed MST (black edge is interconnection, dashed-blue is local connections)}
   \label{fig:mst-exp}
\end{figure}

\begin{figure}[h!]
\captionsetup{justification=centering}
   \begin{subfigure}{0.23\textwidth}
   \centering
       \includegraphics[scale=0.11]{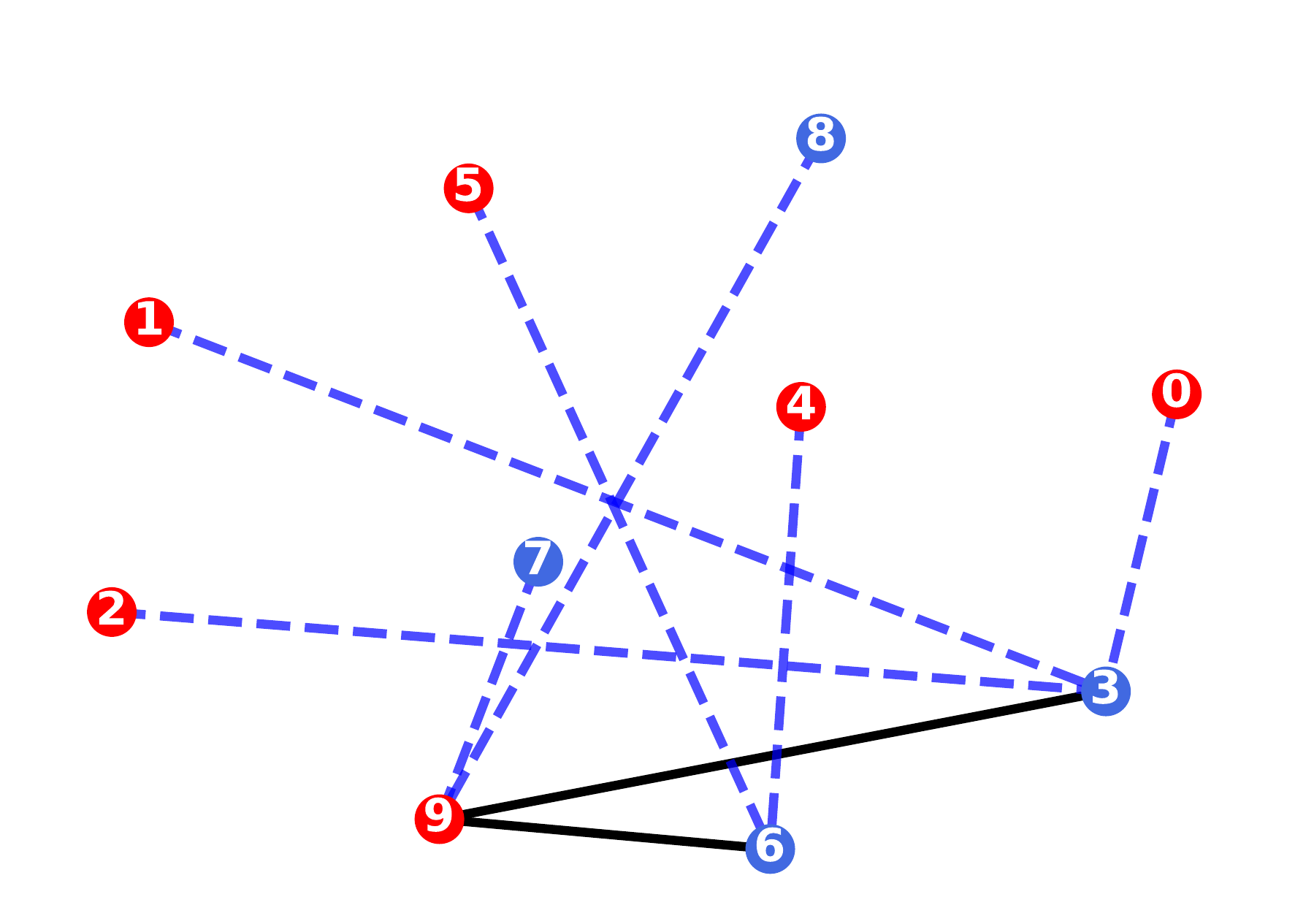} 
       \caption{Complete graph}
       \label{fig:color-exp-complete}
   \end{subfigure}\hfill
   \begin{subfigure}{0.24\textwidth}
   \centering
       \includegraphics[scale=0.11]{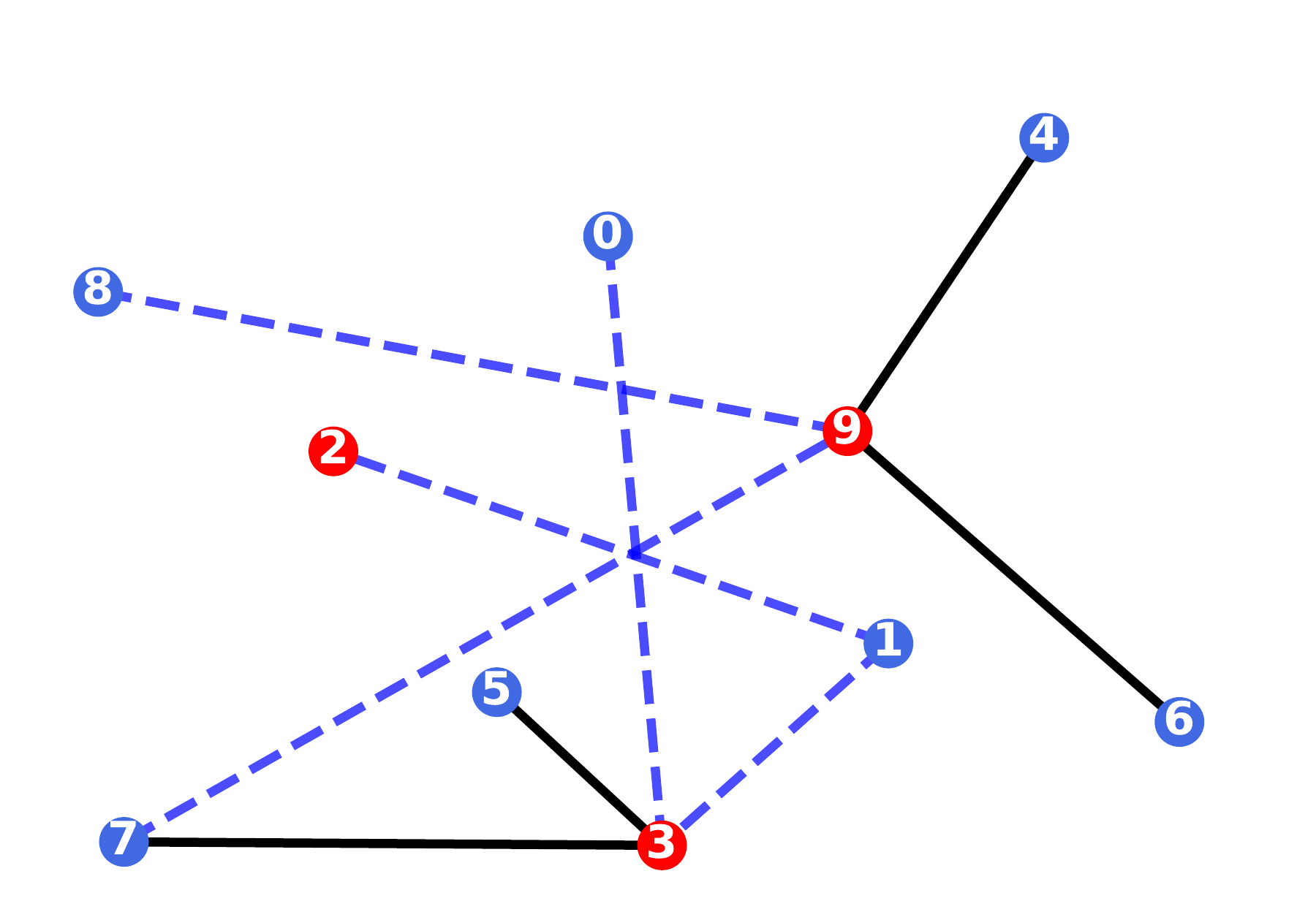}
       \caption{Erdös-Rényi}
       \label{fig:color-exp-renyi}
   \end{subfigure}\hfill
   \begin{subfigure}{0.25\textwidth}
   \centering
       \includegraphics[scale=0.11]{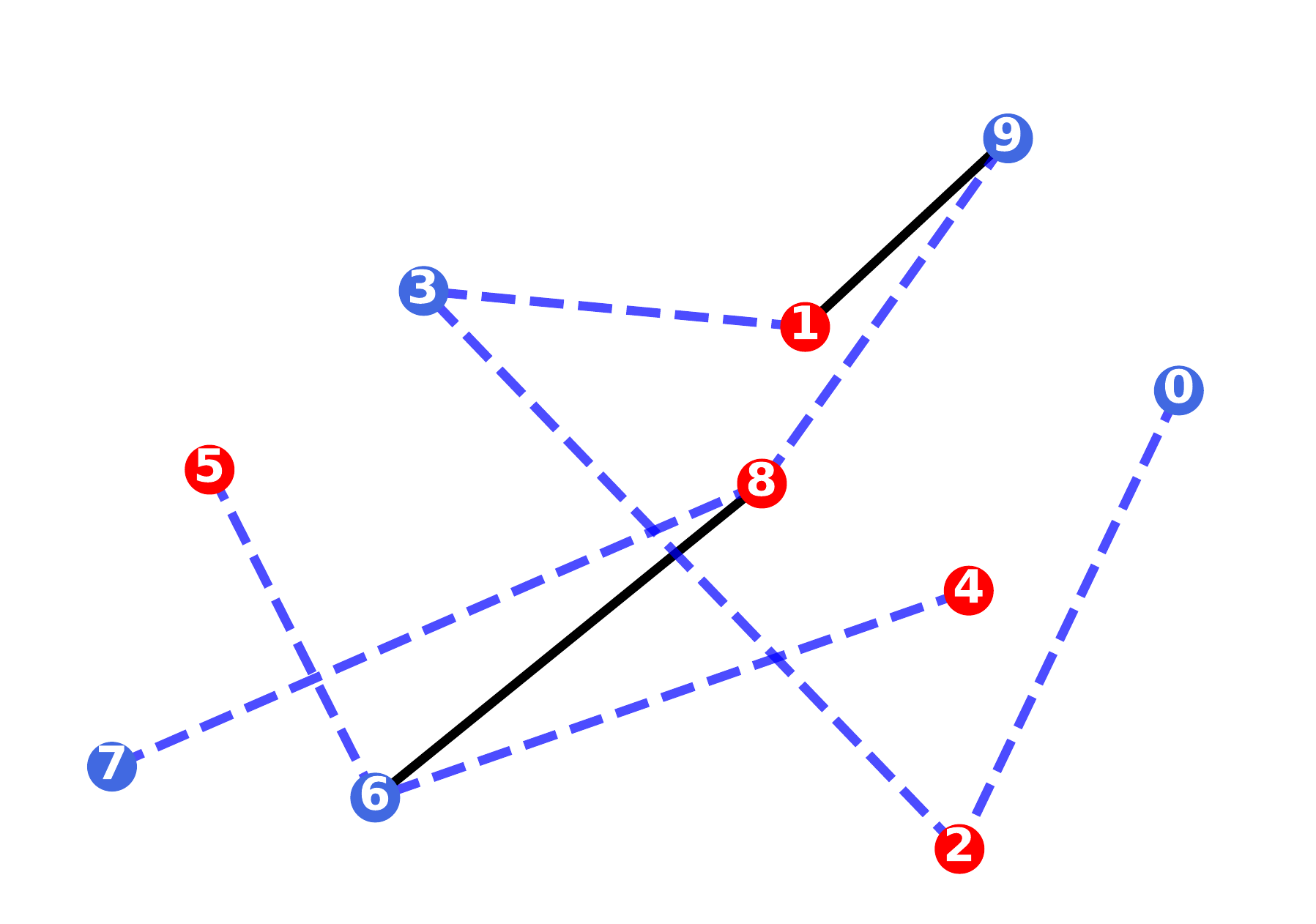}
       \caption{Watts–Strogatz}
       \label{fig:color-exp-watt}
   \end{subfigure}
    \begin{subfigure}{0.23\textwidth}
   \centering
       \includegraphics[scale=0.11]{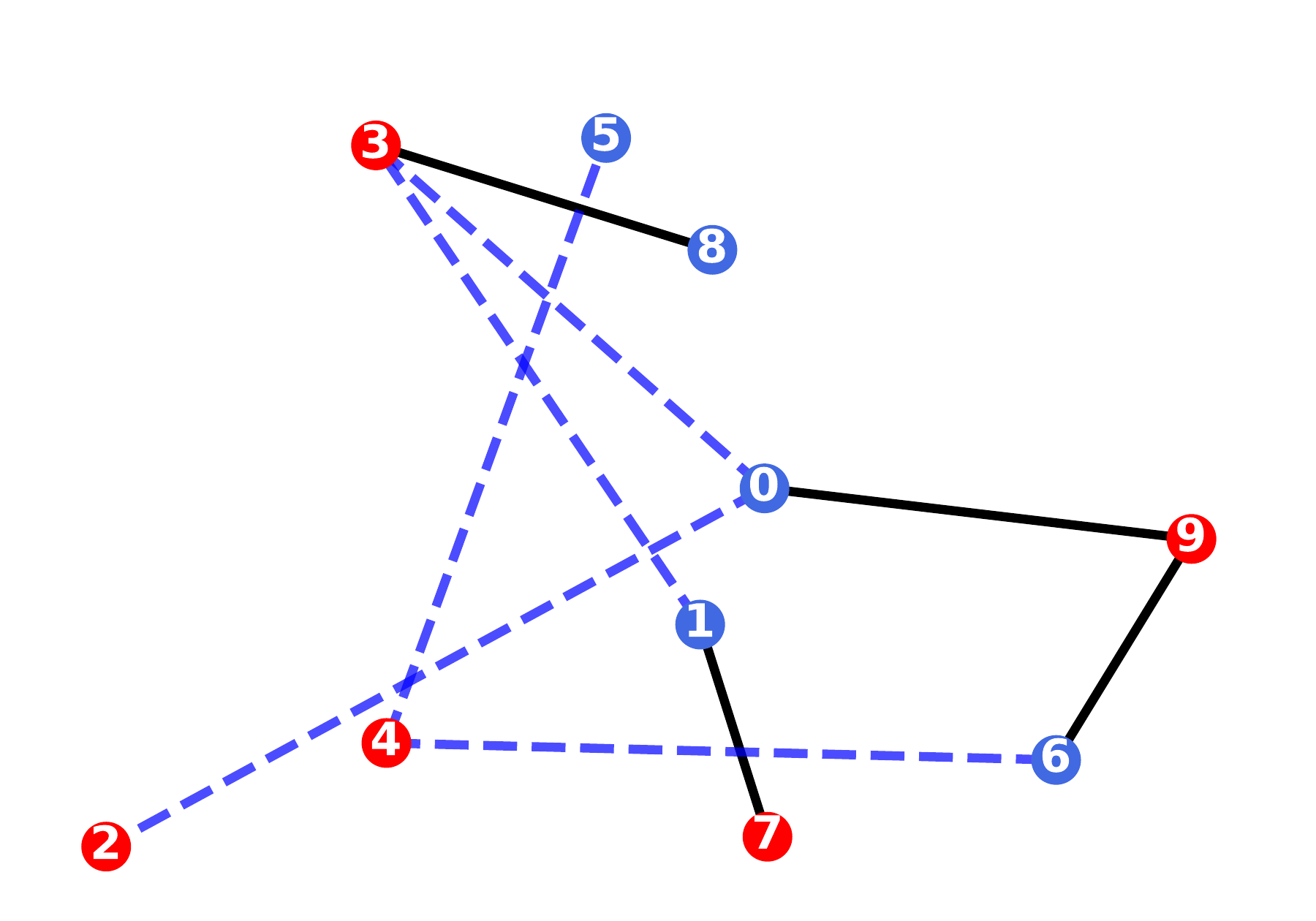}
       \caption{Barabási–Albert}
       \label{fig:color-exp-bara}
   \end{subfigure}
   \caption{Colored MST (black edge is interconnection, dashed-blue is local connections)}
   \label{fig:color-exp}
\end{figure}

All the reported figures in Table \ref{tab:exp-mps}, Table \ref{tab:exp-time} and Table \ref{tab:exp-timetotal} are averaged. Fig. \ref{fig:mst-exp} and Fig. \ref{fig:color-exp} also present the constructed MSTs and coloring results of experimented graphs, respectively. 

It is noted that the report is only about communication, not containing the graph processing time at the moderator nodes. Besides, this work does not include the evaluation of training performance, relying instead on findings from prior studies \cite{beltran2024fedstellar, maejima2024tram} and a practical report by Nguyen et al. \cite{nguyen2024wait}. These sources demonstrated that DFL can maintain comparable accuracy levels to CFL in broadcast or gossip mode.  

\begin{table*}[h!]
\centering
\caption{Bandwidth (MB/s) on different topology and model size}
\label{tab:exp-mps}
\resizebox{\textwidth}{!}{
\def\arraystretch{1.2}
\begin{tabular}{|c|ccllcll|ccccllc|}
\hline
\textbf{} &
  \multicolumn{7}{c|}{\textbf{Broadcast}} &
  \multicolumn{7}{c|}{\textbf{Proposed method}} \\ \hline
\diagbox{Topology}{Model} &
  \multicolumn{1}{c|}{v3s} &
  \multicolumn{1}{c|}{v2} &
  \multicolumn{1}{l|}{b0} &
  \multicolumn{1}{l|}{v3l} &
  \multicolumn{1}{c|}{b1} &
  \multicolumn{1}{l|}{b2} &
  b3 &
  \multicolumn{1}{c|}{v3s} &
  \multicolumn{1}{c|}{v2} &
  \multicolumn{1}{l|}{b0} &
  \multicolumn{1}{l|}{v3l} &
  \multicolumn{1}{c|}{b1} &
  \multicolumn{1}{l|}{b2} &
  \multicolumn{1}{l|}{b3} \\ \hline
Erdös-Rényi &
  \multicolumn{1}{c|}{\multirow{4}{*}{1.785}} &
  \multicolumn{1}{c|}{\multirow{4}{*}{1.096}} &
  \multicolumn{1}{l|}{\multirow{4}{*}{1.011}} &
  \multicolumn{1}{l|}{\multirow{4}{*}{1.066}} &
  \multicolumn{1}{c|}{\multirow{4}{*}{0.842}} &
  \multicolumn{1}{l|}{\multirow{4}{*}{0.839}} &
  \multirow{4}{*}{0.767} &
  \multicolumn{1}{c|}{5.353} &
  \multicolumn{1}{c|}{4.480} &
  \multicolumn{1}{l|}{4.795} &
  \multicolumn{1}{l|}{5.600} &
  \multicolumn{1}{c|}{6.610} &
  \multicolumn{1}{l|}{5.200} &
  \multicolumn{1}{l|}{6.022} \\ \cline{1-1} \cline{9-15} 
Watts–Strogatz &
  \multicolumn{1}{c|}{} &
  \multicolumn{1}{c|}{} &
  \multicolumn{1}{l|}{} &
  \multicolumn{1}{l|}{} &
  \multicolumn{1}{c|}{} &
  \multicolumn{1}{l|}{} &
   &
  \multicolumn{1}{c|}{4.640} &
  \multicolumn{1}{c|}{4.559} &
  \multicolumn{1}{c|}{5.006} &
  \multicolumn{1}{c|}{6.272} &
  \multicolumn{1}{l|}{6.240} &
  \multicolumn{1}{l|}{5.739} &
   6.146\\ \cline{1-1} \cline{9-15} 
Barabási–Albert &
  \multicolumn{1}{c|}{} &
  \multicolumn{1}{c|}{} &
  \multicolumn{1}{l|}{} &
  \multicolumn{1}{l|}{} &
  \multicolumn{1}{c|}{} &
  \multicolumn{1}{l|}{} &
   &
  \multicolumn{1}{c|}{3.969} &
  \multicolumn{1}{c|}{3.600} &
  \multicolumn{1}{c|}{4.204} &
  \multicolumn{1}{c|}{4.665} &
  \multicolumn{1}{l|}{5.794} &
  \multicolumn{1}{l|}{4.861} &
  5.522 \\ \cline{1-1} \cline{9-15} 
Complete &
  \multicolumn{1}{c|}{} &
  \multicolumn{1}{c|}{} &
  \multicolumn{1}{l|}{} &
  \multicolumn{1}{l|}{} &
  \multicolumn{1}{c|}{} &
  \multicolumn{1}{l|}{} &
   &
  \multicolumn{1}{c|}{4.349} &
  \multicolumn{1}{c|}{4.345} &
  \multicolumn{1}{c|}{4.312} &
  \multicolumn{1}{c|}{4.909} &
  \multicolumn{1}{l|}{3.863} &
  \multicolumn{1}{l|}{3.815} &
  4.610 \\ \hline
\end{tabular}
}
\end{table*}

\begin{table*}[h!]
\centering
\caption{Average time (s) for one transfer on different topology and model size}
\label{tab:exp-time}
\resizebox{\textwidth}{!}{
\def\arraystretch{1.2}
\begin{tabular}{|c|ccllcll|ccccllc|}
\hline
\textbf{} &
  \multicolumn{7}{c|}{\textbf{Broadcast}} &
  \multicolumn{7}{c|}{\textbf{Proposed method}} \\ \hline
\diagbox{Topology}{Model} &
  \multicolumn{1}{c|}{v3s} &
  \multicolumn{1}{c|}{v2} &
  \multicolumn{1}{l|}{b0} &
  \multicolumn{1}{l|}{v3l} &
  \multicolumn{1}{c|}{b1} &
  \multicolumn{1}{l|}{b2} &
  b3 &
  \multicolumn{1}{c|}{v3s} &
  \multicolumn{1}{c|}{v2} &
  \multicolumn{1}{l|}{b0} &
  \multicolumn{1}{l|}{v3l} &
  \multicolumn{1}{c|}{b1} &
  \multicolumn{1}{l|}{b2} &
  \multicolumn{1}{l|}{b3} \\ \hline
Erdös-Rényi &
  \multicolumn{1}{c|}{\multirow{4}{*}{6.500}} &
  \multicolumn{1}{c|}{\multirow{4}{*}{12.773}} &
  \multicolumn{1}{l|}{\multirow{4}{*}{20.970}} &
  \multicolumn{1}{l|}{\multirow{4}{*}{20.255}} &
  \multicolumn{1}{c|}{\multirow{4}{*}{37.060}} &
  \multicolumn{1}{l|}{\multirow{4}{*}{42.864}} &
  \multirow{4}{*}{62.576} &
  \multicolumn{1}{c|}{2.167} &
  \multicolumn{1}{c|}{3.125} &
  \multicolumn{1}{l|}{4.421} &
  \multicolumn{1}{l|}{3.857} &
  \multicolumn{1}{c|}{4.720} &
  \multicolumn{1}{l|}{7.077} &
  \multicolumn{1}{l|}{7.971} \\ \cline{1-1} \cline{9-15} 
Watts–Strogatz &
  \multicolumn{1}{c|}{} &
  \multicolumn{1}{c|}{} &
  \multicolumn{1}{l|}{} &
  \multicolumn{1}{l|}{} &
  \multicolumn{1}{c|}{} &
  \multicolumn{1}{l|}{} &
   &
  \multicolumn{1}{c|}{2.500} &
  \multicolumn{1}{c|}{3.071} &
  \multicolumn{1}{c|}{4.235} &
  \multicolumn{1}{c|}{3.444} &
  \multicolumn{1}{l|}{5.000} &
  \multicolumn{1}{l|}{6.412} &
   7.810\\ \cline{1-1} \cline{9-15} 
Barabási–Albert &
  \multicolumn{1}{c|}{} &
  \multicolumn{1}{c|}{} &
  \multicolumn{1}{l|}{} &
  \multicolumn{1}{l|}{} &
  \multicolumn{1}{c|}{} &
  \multicolumn{1}{l|}{} &
   & 
  \multicolumn{1}{c|}{2.923} &
  \multicolumn{1}{c|}{3.888} &
  \multicolumn{1}{c|}{5.042} &
  \multicolumn{1}{c|}{4.630} &
  \multicolumn{1}{l|}{5.385} &
  \multicolumn{1}{l|}{7.571} &
  8.692 \\ \cline{1-1} \cline{9-15} 
Complete &
  \multicolumn{1}{c|}{} &
  \multicolumn{1}{c|}{} &
  \multicolumn{1}{l|}{} &
  \multicolumn{1}{l|}{} &
  \multicolumn{1}{c|}{} &
  \multicolumn{1}{l|}{} &
   &
  \multicolumn{1}{c|}{2.667} &
  \multicolumn{1}{c|}{3.222} &
  \multicolumn{1}{c|}{4.917} &
  \multicolumn{1}{c|}{4.400} &
  \multicolumn{1}{l|}{8.077} &
  \multicolumn{1}{l|}{9.647} &
   10.412\\ \hline
\end{tabular}
}
\end{table*}

\begin{table*}[h!]
\centering
\caption{Average total time (s) to complete one FL communication round}
\label{tab:exp-timetotal}
\resizebox{\textwidth}{!}{
\def\arraystretch{1.12}
\begin{tabular}{|c|ccllcll|ccccllc|}
\hline
\textbf{} &
  \multicolumn{7}{c|}{\textbf{Broadcast}} &
  \multicolumn{7}{c|}{\textbf{Proposed method}} \\ \hline
\diagbox{Topology}{Model} &
  \multicolumn{1}{c|}{v3s} &
  \multicolumn{1}{c|}{v2} &
  \multicolumn{1}{l|}{b0} &
  \multicolumn{1}{l|}{v3l} &
  \multicolumn{1}{c|}{b1} &
  \multicolumn{1}{l|}{b2} &
  b3 &
  \multicolumn{1}{c|}{v3s} &
  \multicolumn{1}{c|}{v2} &
  \multicolumn{1}{l|}{b0} &
  \multicolumn{1}{l|}{v3l} &
  \multicolumn{1}{c|}{b1} &
  \multicolumn{1}{l|}{b2} &
  \multicolumn{1}{l|}{b3} \\ \hline
Erdös-Rényi &
  \multicolumn{1}{c|}{\multirow{4}{*}{10}} &
  \multicolumn{1}{c|}{\multirow{4}{*}{24}} &
  \multicolumn{1}{l|}{\multirow{4}{*}{30}} &
  \multicolumn{1}{l|}{\multirow{4}{*}{30}} &
  \multicolumn{1}{c|}{\multirow{4}{*}{55}} &
  \multicolumn{1}{l|}{\multirow{4}{*}{61}} &
  \multirow{4}{*}{83} &
  \multicolumn{1}{c|}{5.875} &
  \multicolumn{1}{c|}{6.714} &
  \multicolumn{1}{l|}{10.625} &
  \multicolumn{1}{l|}{15.125} &
  \multicolumn{1}{c|}{15.333} &
  \multicolumn{1}{l|}{29} &
  \multicolumn{1}{l|}{33.875} \\ \cline{1-1} \cline{9-15} 
Watts–Strogatz &
  \multicolumn{1}{c|}{} &
  \multicolumn{1}{c|}{} &
  \multicolumn{1}{l|}{} &
  \multicolumn{1}{l|}{} &
  \multicolumn{1}{c|}{} &
  \multicolumn{1}{l|}{} &
   &
  \multicolumn{1}{c|}{3.75} &
  \multicolumn{1}{c|}{5.857} &
  \multicolumn{1}{c|}{10} &
  \multicolumn{1}{c|}{10.333} &
  \multicolumn{1}{l|}{12.571} &
  \multicolumn{1}{l|}{27.75} &
   29.75\\ \cline{1-1} \cline{9-15} 
Barabási–Albert &
  \multicolumn{1}{c|}{} &
  \multicolumn{1}{c|}{} &
  \multicolumn{1}{l|}{} &
  \multicolumn{1}{l|}{} &
  \multicolumn{1}{c|}{} &
  \multicolumn{1}{l|}{} &
   & 
  \multicolumn{1}{c|}{6.5} &
  \multicolumn{1}{c|}{8.2} &
  \multicolumn{1}{c|}{14.2} &
  \multicolumn{1}{c|}{17.125} &
  \multicolumn{1}{l|}{17.5} &
  \multicolumn{1}{l|}{36} &
  38 \\ \cline{1-1} \cline{9-15} 
Complete &
  \multicolumn{1}{c|}{} &
  \multicolumn{1}{c|}{} &
  \multicolumn{1}{l|}{} &
  \multicolumn{1}{l|}{} &
  \multicolumn{1}{c|}{} &
  \multicolumn{1}{l|}{} &
   &
  \multicolumn{1}{c|}{3.16} &
  \multicolumn{1}{c|}{6} &
  \multicolumn{1}{c|}{7.17} &
  \multicolumn{1}{c|}{12.5} &
  \multicolumn{1}{l|}{28.5} &
  \multicolumn{1}{l|}{32.8} &
   35.25\\ \hline
\end{tabular}
}
\end{table*}

\subsection{Impact by model size}
Based on the size of the model, we divide these into three main size categories: small (v2, v3s), medium (b0, v3l), and large (b1, b2, b3) in Table~\ref{tab:model-capacity}. Overall, our method demonstrates remarkable efficiency across all experimented model sizes, as evidenced by its performance in reducing communication times, improving transfer speeds, and optimizing bandwidth usage. Some notable findings are listed below:

\textbf{Small-size models: } for model v3s in the Complete topology, the total time reduces substantially from 10 seconds (broadcast) to 3.16 seconds (our method), demonstrating a nearly 70\% reduction (3.16 times), referred to Table \ref{tab:exp-timetotal}. This is also reflected via the fast single transfer speed of 2.667 seconds and 2.44 times increased bandwidth (Table~\ref{tab:exp-mps}). Transferring these models can benefit from their small size, resulting in significantly reduced communication times and improved transfer speeds. It can be seen that the bandwidth increases from 2.22 (Barabási-v3s) to 4.16 (Watt-v2) and the total transfer time saves up to 4.1 times (Watt-v2) in Table~\ref{tab:exp-timetotal}.

\textbf{Medium-size models: } compared to the small-size models, transferring medium-size models still has some increase in bandwidth from 4.16 (Barabási-b0) to 5.88 (Watt-v3l) (Table~\ref{tab:exp-mps}); however, the total transfer time saves only a bit higher at 4.18 times (Complete-b0) in Table~\ref{tab:exp-timetotal}.

\textbf{Large-size models: } as the model size increases, the enhanced efficiency of our proposed method becomes more pronounced. In particular, bandwidth increases from 4.55 (Complete-b2) to 8.01 (Watt-b3) (Table~\ref{tab:exp-mps}) and the total transfer time saves up to 4.38 times (Watt-b1) (Table~\ref{tab:exp-timetotal}).

\subsection{Impact by topology}
In evaluating our proposed method across the Erdös-Rényi, Watts–Strogatz, Barabási–Albert, and complete topologies, it becomes clear that the method uniformly enhances both bandwidth utilization and completion time for communication rounds, a benefit that is especially prominent when handling larger models. This improvement across diverse network structures, ranging from highly connected to less connected topologies, underscores the method's adaptability and efficiency.

\textbf{Erdös-Rényi:} shows the most substantial improvements, particularly for large models like b3, where the total communication time is reduced significantly (Table~\ref{tab:exp-timetotal}). In particular, the bandwidth efficiency in this topology is remarkable in the increase from 0.767 MB/s to 6.022 MB/s (Table~\ref{tab:exp-mps}).

\textbf{Watt-Strogatz:} demonstrates consistent efficiency improvements, especially in transfer times and bandwidth utilization. These considerable enhancements are witnessed via transferring medium and large models.

\textbf{Barabási-Albert:} can be considered to be the second slowest, after complete for large models. 

\textbf{Complete:} shows the significant improvement in bandwidth utilization (Table~\ref{tab:exp-mps}), in terms of reduced communication round times across small to medium model sizes.

Finally, it is notable that the total time for completing a communication round does not directly correspond to the average time per transfer multiplied by the number of transfers/sending. This discrepancy is largely due to the distances between nodes in the network. To be more specific, some nodes are in proximity (e.g., in the same local network, only need local connection), facilitating rapid data transfer, while others are more distant (e.g., those that need interconnection), resulting in significantly longer transfer times. This variability, which can range from 10 to 60 times depending on node proximity and network conditions, is a critical factor influencing the overall communication efficiency.

Beyond the aforementioned, this comparative analysis reveals significant improvements in the proposed method's efficiency. It demonstrates a substantial reduction in communication round time and transmission time, coupled with enhanced bandwidth, compared to conventional broadcast mechanisms. 

\section{Conclusion}
\label{sec:conclusion}

This work aims to study the correlation between model size and performance latency in DFL to enhance cross-silo applications. To this end, we posed the research question: ``\textit{How does the size of models affect communication latency and efficiency, and how can these factors be optimized to enhance overall DFL system performance?}''  
Via experimental results, we find that the size of models significantly impacts communication latency and efficiency in DFL systems, and optimizing these factors is crucial for enhancing overall performance. 

% Understanding the ratio between model size and performance latency in DFL, 
We proposed an efficient communication method for decentralized collaboration across individual silos, leveraging graph-based algorithms such as MST and graph coloring to restructure the network and minimize concurrent communication. Our approach ensures efficient data flow and congestion-free networks. Unlike prior studies that rely on simulations and idealized environments, which are unreliable in their applicability in practice. Our assessment emphasizes real-world applicability, with evaluations conducted in an ad-hoc network infrastructure setup incorporating actual subnet divisions implemented on physical routers and devices. Besides, all setup selections within this work were thoroughly considered on real-world relevance and reflection. 

Experimental results underscore the comprehensive efficiency of our method across various network topologies, particularly in handling larger models. 
Small models like v3s achieve the greatest efficiency gains, with up to 70\% reduction in communication time and 2.44x bandwidth improvement. Medium models see moderate bandwidth increases but smaller gains in transfer times, while large models benefit significantly, with bandwidth rising from 4.55 to 8.01 MB/s and up to 4.38x faster transfers.
Furthermore, topology plays a pivotal role in determining communication efficiency. The proposed method shows uniform enhancements across various topologies, with the Erdös-Rényi network yielding the most improvements for large models, while the complete topology is superior in bandwidth utilization for small and medium-sized models. 
This consistent performance highlights our method as a potentially scalable and applicable solution in DFL, where maintaining quick update cycles is vital.

\section*{Acknowledgment}
This research is supported by the Tauno Tönning Foundation for Huong Nguyen, by Research Council of Finland through the 6G Flagship program (Grant 318927), and by Business Finland through the Neural pub/sub research project (diary number 8754/31/2022).

\bibliographystyle{IEEEtran}
\bibliography{main}

\vspace{12pt}
\end{document}